\documentclass[,superscriptaddress,twocolumn,preprintnumbers,amssymb,nofootinbib]{revtex4-1}
\usepackage{amsmath, amsthm, amssymb} 

\usepackage{graphics,bm}
\usepackage{epsfig}
\usepackage{graphicx}
\usepackage{amsmath}
\usepackage{wrapfig}
\usepackage{color}
\usepackage{tikz}
\usetikzlibrary{arrows,shapes}
\usetikzlibrary{trees}
\usetikzlibrary{matrix,arrows} 			
\usetikzlibrary{positioning}				
\usetikzlibrary{calc,through}				
\usepackage{pgffor}							

\usetikzlibrary{decorations.pathmorphing}	
\usetikzlibrary{decorations.markings}
\tikzset{
    sigmaCT/.style={draw=black, postaction={decorate},
        decoration={markings,mark=at position .99 with {\arrow[draw=black]{>}},mark=at 		 position .99 with {\arrow[
draw=black]{<}}}},
    pionCT/.style={dashed,draw=black, postaction={decorate},
        decoration={markings,mark=at position .99 with {\arrow[draw=black]{>}},mark=at position .99 with {\arrow[draw=black]{<}}}},    
    fermionCT/.style={draw=black, postaction={decorate},
        decoration={markings,mark=at position .5 with {\arrow[draw=black]{>}},mark=at position .99 with {\arrow[draw=black]{>}},mark=at position .99 with {\arrow[draw=black]{<}}}},    
    fermion/.style={draw=black, postaction={decorate},
        decoration={markings,mark=at position .55 with {\arrow[draw=black]{>}}}},
    fermionbar/.style={draw=black, postaction={decorate},
        decoration={markings,mark=at position .55 with {\arrow[draw=black]{<}}}},
    pion/.style={dashed,draw=black, postaction={decorate}},
    sigma/.style={draw=black, postaction={decorate}}
}

\newcommand{\beq}{\begin{equation}}
\newcommand{\eeq}{\end{equation}}
\newcommand{\bqa}{\begin{eqnarray}}
\newcommand{\eqa}{\end{eqnarray}}

\newcommand{\ms}{\overline{\text{\tiny MS}}}

\def\square{\vcenter{\vbox{\hrule height.4pt
          \hbox{\vrule width.4pt height4pt
          \kern4pt\vrule width.3pt}\hrule height.4pt}}}

\begin{document}

\title{Chiral-density wave versus pion condensation at finite density
and zero temperature}

\author{Jens O. Andersen}
\email{andersen@tf.phys.ntnu.no}
\affiliation{Department of Physics, Faculty of Natural Sciences, NTNU, 
Norwegian University of Science and Technology, H{\o}gskoleringen 5,
N-7491 Trondheim, Norway}
\affiliation{Niels Bohr International Academy, 
Blegdamsvej 17, Copenhagen 2100, Denmark}
\author{Patrick Kneschke}
\email{patrick.kneschke@uis.no}
\affiliation{Faculty of Science and Technology, University of Stavanger,
N-4036 Stavanger, Norway}
\date{\today}

\begin{abstract}
The quark-meson model is often used as a low-energy 
effective model for QCD to 
study the chiral transition at finite temperature $T$, baryon chemical 
potential $\mu_B$, and isospin chemical potential $\mu_I$.
We determine the parameters of the model 
by matching the meson and quark masses, as well as
the pion decay constant to their physical values using the on-shell 
(OS) and 
modified minimal subtraction ($\overline{\rm MS}$)
schemes. In this paper, the 
existence of different phases at zero temperature
is studied. In particular, we 
investigate the competition between an inhomogeneous chiral condensate
and a homogeneous 
pion condensate. For the inhomogeneity, we use a chiral-density
wave ansatz. For a sigma mass of $600$ MeV, we find
that an inhomogeneous chiral condensate exists
only for pion masses below approximately 37 MeV. 
We also show that due to our parameter fixing,
the onset of pion condensation takes place exactly at 
$\mu_I^c={1\over2}m_{\pi}$
in accordance with exact results.
\end{abstract}
\keywords{Dense QCD,
chiral transition, }

\maketitle

\section{Introduction}
The phases of dense QCD as functions of the baryon chemical potential
$\mu_B$ or the quark chemical potential $\mu={1\over3}\mu_B$
have been studied in detail since the first
phase diagram was suggested in the 1970s \cite{raja,alford,fukurev}.
For baryon chemical potentials 
lower than the nucleon mass $m_N$ and at $T=0$, we
are in the vacuum phase. 
For larger values of the baryon chemical potential, one expected 
a first-order transition to quark matter, a phase in which 
chiral symmetry is approximately restored and where quarks are no
longer confined to the individual nucleons.
It turns out that the phase diagram is very rich. For asymptotically
large values of the quark chemical potential, we have one of the
few rigorous results. Due to asymptotic freedom, the system is
weakly interacting and an attractive quark-quark interaction from 
one-gluon exchange gives rise to color superconductivity and the
so called color-flavor locked phase.
For lower values of the quark chemical potential, one cannot
use perturbative QCD  and instead one has to use
low-energy effective models of QCD such as the 
quark-meson (QM) model or 
Nambu-Jona-Lasinio (NJL)
model. These models predict a plethora of superconducting phases depending
on e.g. the quark masses and other external parameters such as
magnetic fields \cite{ferrer1,harmen, nor1}.
Other possibilities include the quarkyonic phase \cite{rob1,rob2}, whose
existence is unambiguously argued in the large-$N_c$ limit, where
$N_c$ is the number of colors.

Another interesting aspect of the QCD phase diagram at $T=0$ is the
possibility of inhomogeneous phases, see  \cite{buballarev, casal}
for reviews.
These are phases where two quarks (or a quark and an antiquark)
with  momenta ${\bf p}+{\bf q}$ and $-{\bf p}+{\bf q}$
pair, resulting in a Cooper pair with net momentum $2{\bf q}$
and a complex order parameter
of the form $\Delta e^{2i{\bf q}\cdot{\bf x}}$.

Instead of considering 
only the baryon chemical potential, one can allow for
an independent chemical potential $\mu_f$
for each quark flavor $f$. For two quark flavors, using $\mu_u$
and $\mu_d$ is equivalent to using $\mu_B$ and an isospin chemical
potential $\mu_I$. Systems at zero baryon chemical potential and finite
isospin chemical potential are of particular interest, since the
fermion determinant is real and one can perform lattice simulations
using standard importance sampling techniques, see
e.g. \cite{kogut1,kogut2,gergy1,gergy2,gergy3}.
The picture that emerges from the simulations is that at $T=0$, the 
chiral condensate is constant for $\mu_I$ below a critical
value, 
$\mu_I^c={1\over2}m_{\pi}$.\footnote{Depending on convention, 
$\mu_I^c=m_{\pi}$ is also frequently found in the literature.}
The critical value of 
the isospin chemical potential marks the onset of pion condensation and 
the transition is of second order. This picture is consistent with
the predictions of chiral perturbation theory \cite{son,kim}.

In addition to chiral perturbation 
theory \cite{son,kim,loewe,fragaiso,carigchpt},
there have also been a number of other approaches and model 
calculations studying various
aspects of the QCD phase diagram at finite isospin density.
These include the resonance gas model \cite{restoublan}, 
random matrix models \cite{random}, the
NJL model \cite{2fbuballa,toublannjl,bar2f,he2f,heman2,heman,ebert1,ebert2,lars,2fabuki,heman3,he3f}, 
the quark-meson model \cite{lorenz,ueda,qmstiele}, perturbative
QCD \cite{graf}, and hard-thermal-loop perturbation theory \cite{htlpt}.

In the present paper, we use the quark-meson model to study possible
phases at $T=0$ and at finite baryon and isospin chemical potentials. 
In particular, we extend certain aspects of earlier
studies \cite{nakano,nick1,bubsc,balli}
by looking at the competition between an inhomogeneous chiral
condensate and a homogeneous pion condensate.
Studies of the competition between homogeneous and inhomogeneous condensates
have been carried out in the 1+1 dimensional NJL model in 
\cite{1p1klim,jens1+1,khunja}.

The paper is organized as follows. 
In Sec. II, we briefly discuss the quark-meson model and in Sec. III we 
calculate the effective potential in the mean-field approximation.
In Sec. IV, we present the phase diagram in the $\mu$--$\mu_I$ plane
at zero temperature and in Sec. V, we summarize and conclude.
In Appendix A, we list a number of integrals needed in the calculations, while
Appendix B provides the reader with some details of how the parameters
are determined. In Appendix $C$, we show that the critical
isospin chemical potential is exactly $\mu_I^c={1\over2}m_{\pi}$
in our approximation.

\section{Quark-meson model}
The Lagrangian of the two-flavor quark-meson model 
in Minkowski space is 
\bqa\nonumber
{\cal L}&=&
{1\over2}\left[(\partial_{\mu}\sigma)(\partial^{\mu}\sigma)
+(\partial_{\mu} \pi_3)(\partial^{\mu} \pi_3)
\right]
\\&&\nonumber
+(\partial_{\mu}+2i\mu_I\delta_{\mu}^0)\pi^+(\partial^{\mu}-2i\mu_I\delta_{0}^{\mu})
\pi^-
\\&&\nonumber-{1\over2}m^2(\sigma^2+\pi_3^2+2\pi^+\pi^-)
-{\lambda\over24}(\sigma^2+\pi_3^2+2\pi^+\pi^-)^2
\\ && \nonumber
+h\sigma+\bar{\psi}\left[
i/\!\!\!\partial
+\mu_f
\gamma^0
-g(\sigma+i\gamma^5{\boldsymbol\tau}\cdot{\boldsymbol\pi})\right]\psi\;,
\\ &&
\label{lag}
\eqa
where $\psi$ is 
a color $N_c$-plet, a four-component Dirac spinor as well as a flavor doublet 
\bqa
\psi&=&
\left(
\begin{array}{c}
u\\
d
\end{array}\right)\;,
\eqa
and $\mu_f={\rm diag}(\mu_u,\mu_d)$,
where $\mu_u$ and $\mu_d$, are the quark chemical potentials,
$\mu_I$ is the isospin chemical potential,
$\tau_i$ ($i=1,2,3$) are the Pauli matrices in flavor space, 
${\boldsymbol\pi}=(\pi_{1},\pi_{2},\pi_{3})$, and
$\pi^{\pm}={1\over\sqrt{2}}(\pi_1\pm i\pi_2)$.

Apart from the global $SU(N_c)$ symmetry, 
the Lagrangian~(\ref{lag}) 
has a 
$U(1)_B\times SU(2)_L\times SU(2)_R$ symmetry for 
$h=0$ and a $U(1)_B\times SU(2)_V$ symmetry
for $h\neq0$. 
When $\mu_u\neq\mu_d$, this symmetry is reduced to 
$U(1)_B\times U_{I_3L}(1)\times U_{I_3R}(1)$ for $h=0$ and
$U(1)_B\times U_{I_3}(1)$ for $h\neq0$. 

The number density associated with a chemical potential $\mu_A$ is
\bqa
n_A&=&-{\partial V\over\partial\mu_A}\;,
\eqa
where $V$ is the effective potential.
The baryon and isospin densities can be expressed in terms of the
quark densities $n_u$ and $n_d$ as
\bqa
\label{nb}
n_B&=&{1\over3}(n_u+n_d)\;,
\\
n_I&=&n_u-n_d\;.
\label{ni}
\eqa
Eqs. (\ref{nb})--(\ref{ni}) together with the chain rule
can be used to derive relations among the
baryon and isospin chemical potentials and the quark chemical potentials.
We have
\bqa\nonumber
n_I&=&-{\partial V\over\partial\mu_I}
\\ \nonumber
&=&-\left({\partial V\over\partial\mu_u}-{\partial V\over\partial\mu_d}\right)
\\
&=&-\left({\partial\mu_u\over\partial\mu_I}{\partial V\over\partial\mu_u}
+{\partial\mu_d\over\partial\mu_I}{\partial V\over\partial\mu_d}\right)
\;.
\eqa
This yields
\bqa
{\partial\mu_u\over\partial\mu_I}
=-{\partial\mu_u\over\partial\mu_I}=1\;.
\eqa
Similarly, we find ${\partial\mu_u\over\partial\mu_B}={\partial\mu_d\over\partial\mu_B}={1\over3}$.
From this, we find the following relations among the chemical potentials
\bqa
\label{rr}
\mu_u&=&{1\over3}\mu_B+\mu_I\;,\\
\mu_d&=&{1\over3}\mu_B-\mu_I\;.
\label{rrr}
\eqa
Introducing the quark chemical potential $\mu={1\over3}\mu_B$ and 
inverting the
relations (\ref{rr})--(\ref{rrr}), we find
\bqa
\mu&=&{1\over2}(\mu_u+\mu_d)\;,\\
\mu_I&=&{1\over2}(\mu_u-\mu_d)\;.
\eqa
In the following, we will express the equations and our result
in terms of 
$\mu$ and $\mu_I$ instead of $\mu_u$ and $\mu_d$.

\section{Effective potential}
In the following, we allow for an inhomogeneous chiral condensate. To be 
specific, we consider a one-dimensional
chiral-density wave with a wave vector ${\bf q}$ pointing in the 
positive $z$-direction.
The expectation values of the fields are written as
\bqa
\sigma&=&\phi_0\cos(qz)\;,\hspace{0.5cm}
\pi_3=\phi_0\sin(qz)
\\
\pi_1&=&\pi_0
\;,\hspace{1.7cm}
\pi_2=0
\;,
\eqa
where $\phi_0$ and $\pi_0$ are constant in space. The latter represents
a homogeneous pion condensate.
A pion condensate breaks the $U_{I_3L}(1)\times U_{I_3R}(1)$
symmetry to $U_{I_3V}(1)$ or the $U_{I_3}(1)$ symmetry.
Introducing 
$\Delta=g\phi_0$
and $\rho=g\pi_0$, 
the tree-level potential in Euclidean space
can be written as 
\bqa\nonumber
V_0&=&{1\over2}{q^2\over g^2}\Delta^2
+{1\over2}{m^2\over g^2}\Delta^2
+{1\over2}{m^2-4\mu_I^2\over g^2}\rho^2
\\ &&
+{\lambda\over24g^4}\left(\Delta^2+\rho^2\right)^2
\label{v0}
-{h\over g}\Delta\cos(qz)\delta_{q,0}\;.
\eqa
The Kronecker-delta $\delta_{q,0}$ in the last term in 
Eq. (\ref{v0}) is necessary since a term $\Delta\cos(qz)$
vanishes for nonzero $q$
upon integrating the free energy density over a sufficiently
large spatial volume.

Expressing the parameters in the Lagrangian in terms of 
the sigma mass $m_{\sigma}$, pion mass $m_{\pi}$, 
pion decay constant $f_{\pi}$, and quark mass $m_q$ , we find
\bqa
\label{rel1}
m^2&=&-{1\over2}\left(m_{\sigma}^2-3m_{\pi}^2\right)\;,\,
\lambda=3{\left(m_{\sigma}^2-m_{\pi}^2\right)\over f_{\pi}^2}\;,
\\
g^2&=&{m_q^2\over f_{\pi}^2}\;,
\hspace{2.1cm}
h=m_{\pi}^2f_{\pi}\;.
\label{rel2}
\eqa
Inserting these relations, we can write the tree-level potential as
\bqa\nonumber
V_0&=&{1\over2}f_{\pi}^2q^2{\Delta^2\over m_q^2}
-{1\over4}f_{\pi}^2(m_{\sigma}^2-3m_{\pi}^2){\Delta^2+\rho^2\over m_q^2}
-
2\mu_I^2f_{\pi}^2{\rho^2\over m_q^2}
\\ \nonumber
&&+{1\over8}f_{\pi}^2(m_{\sigma}^2-m_{\pi}^2){(\Delta^2+\rho^2)^2\over m_q^4}
\\
&&-m_{\pi}^2f_{\pi}^2{\Delta\over m_q}\cos(qz)\delta_{q,0}\;.
\eqa
Since we want to integrate out the fermions, we need the part
of the Lagrangian 
that is bilinear in the quark fields
\bqa\nonumber
{\cal L}_2&=&
\bar{\psi}\bigg[
i\gamma^{\mu}\partial_{\mu}+(\mu+\tau_3\mu_I)\gamma^0
-\Delta e^{i\gamma^5\tau_3qz}
\\ &&
-i\tau_1\gamma^5\rho
\bigg]\psi\;.
\eqa
We next redefine the quark fields, 
$\psi\rightarrow e^{-{1\over2}i\gamma^5\tau_3qx}\psi$
and $\bar{\psi}\rightarrow\bar{\psi}e^{-{1\over2}i\gamma^5\tau_3qx}$.
The transformation of the field $\psi$ amounts to a unitary transformation
of the Dirac Hamiltonian, 
${\cal H}\rightarrow{\cal H}^{\prime}
=e^{{1\over2}i\gamma^5\tau_3qx}{\cal H}e^{-{1\over2}i\gamma^5\tau_3qx}$.
The Dirac operator $D$ then reads
\bqa \nonumber
D&=&\bigg[
i\gamma^{\mu}\partial_{\mu}+(\mu+\tau_3\mu_I)\gamma^0
-\Delta +{1\over2}\gamma^5\gamma^3\tau_3q
\\ &&
-i\tau_1\gamma^5\rho
\bigg]\;.
\eqa
In momentum space, the Dirac operator is
\bqa \nonumber
D&=&\left[/\!\!\!p
+(\mu+\tau_3\mu_I)\gamma^0-\Delta
+{1\over2}\gamma^5\gamma^3\tau_3q
-i\tau_1\gamma^5\rho
\right]\;.
\\
\eqa
\begin{widetext}
The quark energies can be read off from 
the zeros of the determinant of the
Dirac operator, which are found to be, see e.g. \cite{dautry}
\bqa
E_u^{\pm}=E(\pm q,-\mu_I)\;,
\hspace{1cm}
E_d^{\pm}=E(\pm q,\mu_I)\;,
\hspace{1cm}
E_{\bar{u}}^{\pm}=E(\pm q,\mu_I)\;,
\hspace{1cm}E_{\bar{d}}^{\pm}=E(\pm q,-\mu_I)\;.
\eqa
where we have defined
\bqa
E(q,\mu_I)&=&
\left[
\left(\sqrt{p_{\perp}^2+
\left(\sqrt{p_{\parallel}^2+\Delta^2}+{q\over2}\right)^2}+
{\mu_I}\right)^2
+\rho^2\right]^{1\over2}\;,
\eqa
and 
where the wave vector ${\bf q}$ points in the positive $p_{\parallel}$ direction.

The one-loop  contribution to the effective potential is
\bqa
V_1&=&-{1\over2}N_c\int_p\left(E_u^{\pm}+E_d^{\pm}
+E_{\bar{u}}^{\pm}+E_{\bar{d}}^{\pm}\right)\;,
\label{v1}
\eqa
where a sum over $\pm$ is implied and 
the integral is in $d=3-2\epsilon$ dimensions (See Appendix A).
The integral in Eq. (\ref{v1}) 
is ultraviolet divergent and in order to 
isolate the divergences, we need to expand the energies in powers of
$q$ and $\mu_I$ to the appropriate orders. This yields
\bqa\nonumber
V_{\rm div}&=&
-4N_c\int_p\left[
\sqrt{p^2+\Delta^2+\rho^2}
+{\mu_I^2\rho^2\over2({p^2+\Delta^2+\rho^2)^{3\over2}}}
+{3q^2\mu_I^2\rho^2(4\Delta^2+4p_{\parallel}^2-p_{\perp}^2-\rho^2)
\over16(p^2+\Delta^2+\rho^2)^{7\over2}}
\right.\\ &&\left. \nonumber
+{q^2(p_{\perp}^2+\rho^2)\over8({p^2+\Delta^2+\rho^2)^{3\over2}}}
+{q^4(p_{\perp}^2+\rho^2)(4\Delta^2+4p_{\parallel}^2-p_{\perp}^2-\rho^2)
\over128({p^2+\Delta^2+\rho^2)^{7\over2}}}
\right]\\ \nonumber
&=&
{2N_c\over(4\pi)^2}
\left({e^{\gamma_E}\Lambda^2\over\Delta^2+\rho^2}\right)^{\epsilon}
\left\{
2\left(\Delta^2+\rho^2\right)^2
\Gamma(-2+\epsilon)+q^2\Delta^2\Gamma(\epsilon)
-4\mu_I^2\rho^2\Gamma(\epsilon)
-{2q^2\mu_I^2}{\Delta^2\rho^2\over(\Delta^2+\rho^2)^2}\Gamma(2+\epsilon)
\right.\\ &&\left.
-{q^4\over12}{\Delta^2\over(\Delta^2+\rho^2)^2}\left[(1-\epsilon)\Delta^2
+2\rho^2\right]
\Gamma(1+\epsilon)
\right\}\;.
\eqa
The remainder $V_{\rm fin}$ is finite and reads
\bqa
V_{\rm fin}&=&V_1-V_{\rm div}\;.
\eqa
Note that $V_{\rm fin}$ can be evaluated directly in $d=3$ dimensions.
In the case $\rho=\mu_I=0$, one can calculate $V_{\rm fin}$ analytically,
see Ref. \cite{crew} for an explicit evaluation.
In the present case, it must be evaluated numerically.
Using the expressions for the integrals listed in Appendix A, we can
write the unrenormalized one-loop effective potential
$V=V_0+V_1$ as
\bqa\nonumber
V&=&{1\over2}{q^2\over g^2}\Delta^2
+{1\over2}{m^2\over g^2}\Delta^2
+{1\over2}{m^2-4\mu_I^2\over g^2}\rho^2
+{\lambda\over24g^4}(\Delta^2+\rho^2)^2
-{h\over g}\Delta\cos(qz)\delta_{q,0}
\\ \nonumber
&&+{2N_c\over(4\pi)^2}\left(\Lambda^2\over\Delta^2+\rho^2\right)^{\epsilon}
\left[\left(\Delta^2+\rho^2\right)^2\left({1\over\epsilon}+{3\over2}\right)
+q^2\Delta^2{1\over\epsilon}
-4\mu_I^2\rho^2{1\over\epsilon}
-{q^4\over12}{\Delta^2(\Delta^2+2\rho^2)\over(\Delta^2+\rho^2)^2}
\right.\\ &&\left.
-2{q^2\mu_I^2}{\Delta^2\rho^2\over(\Delta^2+\rho^2)^2}
\right]+V_{\rm fin}
+{\cal O}(\epsilon)
\;.
\eqa
The unrenormalized one-loop effective potential 
contains poles in $\epsilon$, which are removed by 
mass and coupling constant renormalization. 
In the $\overline{\rm MS}$ scheme this is achieved by 
making the substitutions $m^2\rightarrow Z_{m^2}m^2$,
$\lambda\rightarrow Z_{\lambda}\lambda$,
$g^2\rightarrow Z_{g^2}g^2$, and 
$h\rightarrow Z_{h}h$, where 
\bqa
Z_{m^2}=1+{4N_cg^2\over(4\pi)^2\epsilon}\;,
\hspace{0.4cm}
Z_{\lambda}=1+{8N_c\over(4\pi)^2\epsilon}\left[
g^2-6{g^4\over\lambda}\right]\;,
\hspace{0.4cm}
Z_{g^2}=1+{4N_cg^2\over(4\pi)^2\epsilon}\;,
\hspace{0.4cm}
Z_{h}=1+{2N_cg^2\over(4\pi)^2\epsilon}\;,
\hspace{0.4cm}
\eqa
The renormalized one-loop effective potential then reads
\bqa\nonumber
V_{\rm 1-loop}&=&{1\over2}{q^2\over g^2_{\ms}}\Delta^2+
{1\over2}{m^2_{\ms}\over g^2_{\ms}}\Delta^2
+{1\over2}{m^2_{\ms}-4\mu_I^2\over g^2_{\ms}}\rho^2
+{\lambda_{\ms}\over24g_{\ms}^4}\left(\Delta^2+\rho^2\right)^2
-{h_{\ms}\over g_{\ms}}\Delta\cos(qz)\delta_{q,0}
\\ \nonumber
&&
+{2N_c\over(4\pi)^2}\Bigg\{
\left[\left(\Delta^2+\rho^2\right)^2+q^2\Delta^2-4\mu_I^2\rho^2\right]
\log\left({\Lambda^2\over\Delta^2+\rho^2}\right)
+{3\over2}\left(\Delta^2+\rho^2\right)^2
-{q^4\over12}{\Delta^2(\Delta^2+2\rho^2)\over(\Delta^2+\rho^2)^2}
\\ &&
-2{q^2\mu_I^2}{\Delta^2\rho^2\over(\Delta^2+\rho^2)^2}
\Bigg\}+V_{\rm fin}\;,
\label{veff}
\eqa
where the subscript ${\overline{\rm MS}}$ indicates that the
parameters are running with the renormalization scale $\Lambda$.
In Appendix B, we discuss how one can express the
parameters in the $\overline{\rm MS}$ scheme in terms
of physical masses and couplings. Using 
Eqs. (\ref{sol1})--(\ref{sol5}), the final expression for the one-loop
effective potential in the large-$N_c$ limit becomes
\bqa\nonumber
V_{\rm1-loop}&=&{1\over2}f_{\pi}^2q^2
\left\{1-\dfrac{4 m_q^2N_c}{(4\pi)^2f_\pi^2}
\left[\log\mbox{$\Delta^2+\rho^2\over m_q^2$}
+F(m_\pi^2)+m_\pi^2F^{\prime}(m_\pi^2)\right]
\right\}{\Delta^2\over m_q^2}
\\ \nonumber &&
+\dfrac{3}{4}m_\pi^2 f_\pi^2
\left\{1-\dfrac{4 m_q^2N_c}{(4\pi)^2f_\pi^2}m_\pi^2F^{\prime}(m_\pi^2)
\right\}\dfrac{\Delta^2+\rho^2}{m_q^2}
\\ \nonumber &&
 -\dfrac{1}{4}m_\sigma^2 f_\pi^2
\left\{
1 +\dfrac{4 m_q^2N_c}{(4\pi)^2f_\pi^2}
\left[ \left(1-\mbox{$4m_q^2\over m_\sigma^2$}
\right)F(m_\sigma^2)
 +\dfrac{4m_q^2}{m_\sigma^2}
-F(m_\pi^2)-m_\pi^2F^{\prime}(m_\pi^2)
\right]\right\}\dfrac{\Delta^2+\rho^2}{m_q^2} 
\\ \nonumber &&
-2\mu_I^2f_\pi^2
\left\{1-\dfrac{4 m_q^2N_c}{(4\pi)^2f_\pi^2}
\left[\log\mbox{$\Delta^2+\rho^2\over m_q^2$}
+F(m_\pi^2)+m_\pi^2F^{\prime}(m_\pi^2)\right]
\right\}{\rho^2\over m_q^2}
\\ \nonumber
 & & + \dfrac{1}{8}m_\sigma^2 f_\pi^2
\left\{ 1 -\dfrac{4 m_q^2  N_c}{(4\pi)^2f_\pi^2}\left[
\dfrac{4m_q^2}{m_\sigma^2}
\left( 
\log\mbox{$\Delta^2+\rho^2\over m_q^2$}
-\mbox{$3\over2$}
\right) -\left( 1 -\mbox{$4m_q^2\over m_\sigma^2$}\right)F(m_\sigma^2)
+F(m_\pi^2)+m_\pi^2F^{\prime}(m_\pi^2)\right]
 \right\}\dfrac{(\Delta^2+\rho^2)^2}{m_q^4}
\\ \nonumber&&
- \dfrac{1}{8}m_\pi^2 f_\pi^2
\left[1-\dfrac{4 m_q^2N_c}{(4\pi)^2f_\pi^2}m_\pi^2F^{\prime}(m_\pi^2)\right]
\dfrac{(\Delta^2+\rho^2)^2}{m_q^4}
-m_\pi^2f_\pi^2\left[
1-\dfrac{4 m_q^2  N_c}{(4\pi)^2f_\pi^2}m_\pi^2F^{\prime}(m_\pi^2)
\right]\dfrac{\Delta}{m_q}\cos(qz)\delta_{q,0}
\\ 
&&
-{N_c\over6(4\pi)^2}\left[
{q^4}{\Delta^2(\Delta^2+2\rho^2)\over(\Delta^2+\rho^2)^2}
+24{q^2\mu_I^2}{\Delta^2\rho^2\over(\Delta^2+\rho^2)^2}
\right]
+V_{\rm fin}\;.
\label{fullb}
\eqa
The effective potential must be independent
of $q$ in the limit $\Delta\rightarrow0$, it cannot depend on the wave vector
if the magnitude of the order parameter is zero.
We have checked numerically
that this is the case for $V_{\rm 1-loop}$ in Eq. (\ref{fullb}). 

The matter part of the one-loop effective potential is
\bqa
V_{\rm mat}&=&-N_cT\int_p\left\{
\log\left[1+e^{-\beta(E_{u}^{\pm}-\mu)}\right]
+\log\left[1+e^{-\beta(E_{d}^{\pm}-\mu)}\right]
+\log\left[1+e^{-\beta(E_{\bar{u}}^{\pm}+\mu)}\right]
+\log\left[1+e^{-\beta(E_{\bar{d}}^{\pm}+\mu)}\right]
\right\}\;.
\eqa
In the limit $T\rightarrow0$, this contribution reads
\bqa
V_{\rm mat}&=&N_c\int_p\left[(E_u^{\pm}-\mu)\theta(\mu-E_u^{\pm})
+(E_d^{\pm}-\mu)\theta(\mu-E_d^{\pm})\right]\;.
\label{fd}
\eqa
\end{widetext}
The complete one-loop effective potential is then the sum of
Eqs. (\ref{fullb}) and (\ref{fd}) and is denoted by
$V_{\rm full}$.

\section{Phase diagram}
In this section, we discuss the phase diagram in the 
$\mu$--$\mu_I$ plane at $T=0$. In the numerical work below, we set
$N_c=3$, $m_{\sigma}=600$ MeV, and $f_{\pi}=93$ MeV. In the chiral limit
$m_{\pi}=0$, while at the physical point, we set $m_{\pi}=140$ MeV.

\subsection{Homogeneous chiral condensate}
We first restrict ourselves to a homogeneous chiral condensate
as well as a homogeneous pion condensate, i.e. we set $q=0$
in Eqs. (\ref{fullb}) and (\ref{fd}). In the remainder of this subsection
we consider only the physical point.

\begin{figure}[htb]
\includegraphics[width=0.4\textwidth]{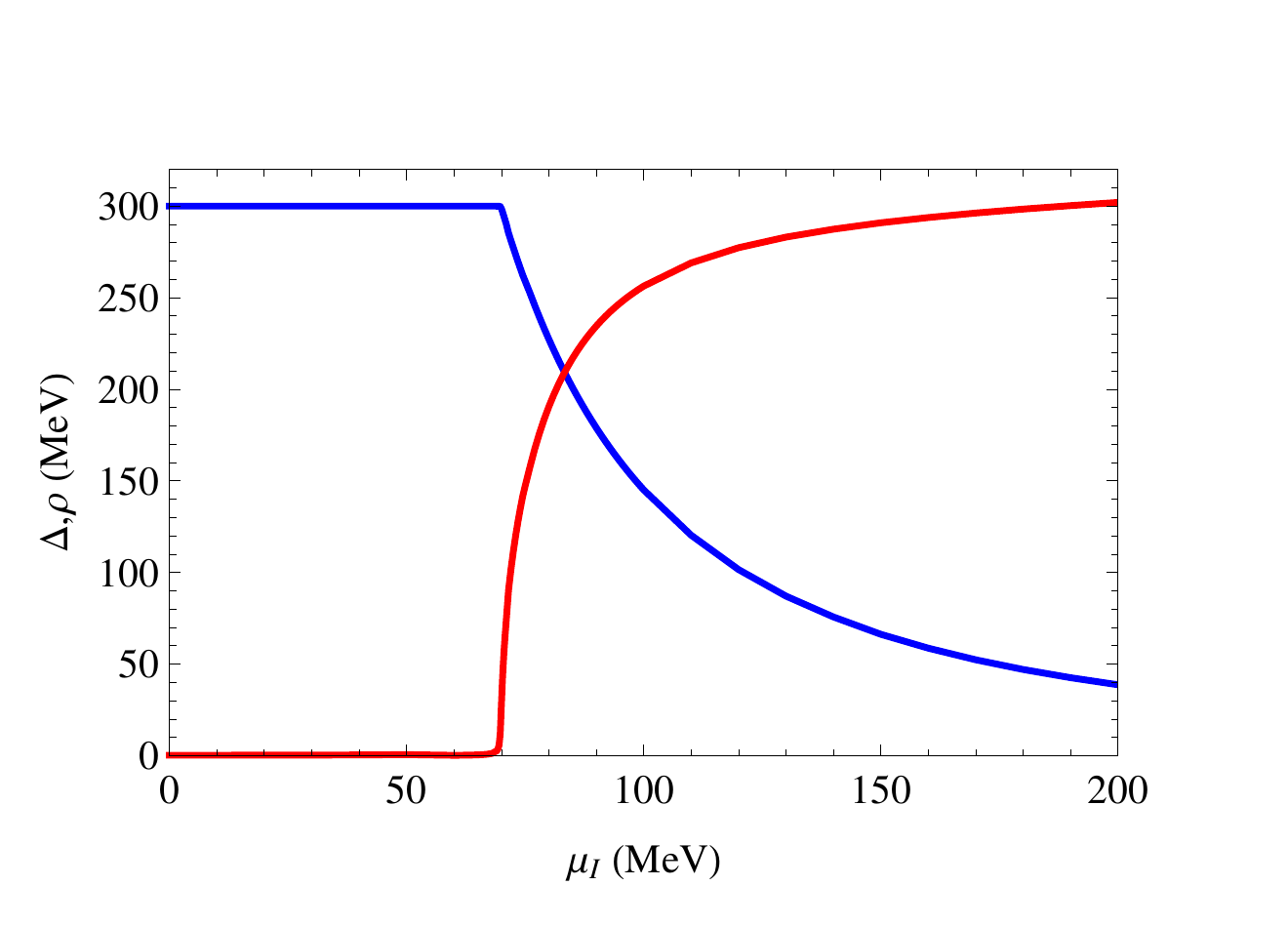}
\includegraphics[width=0.4\textwidth]{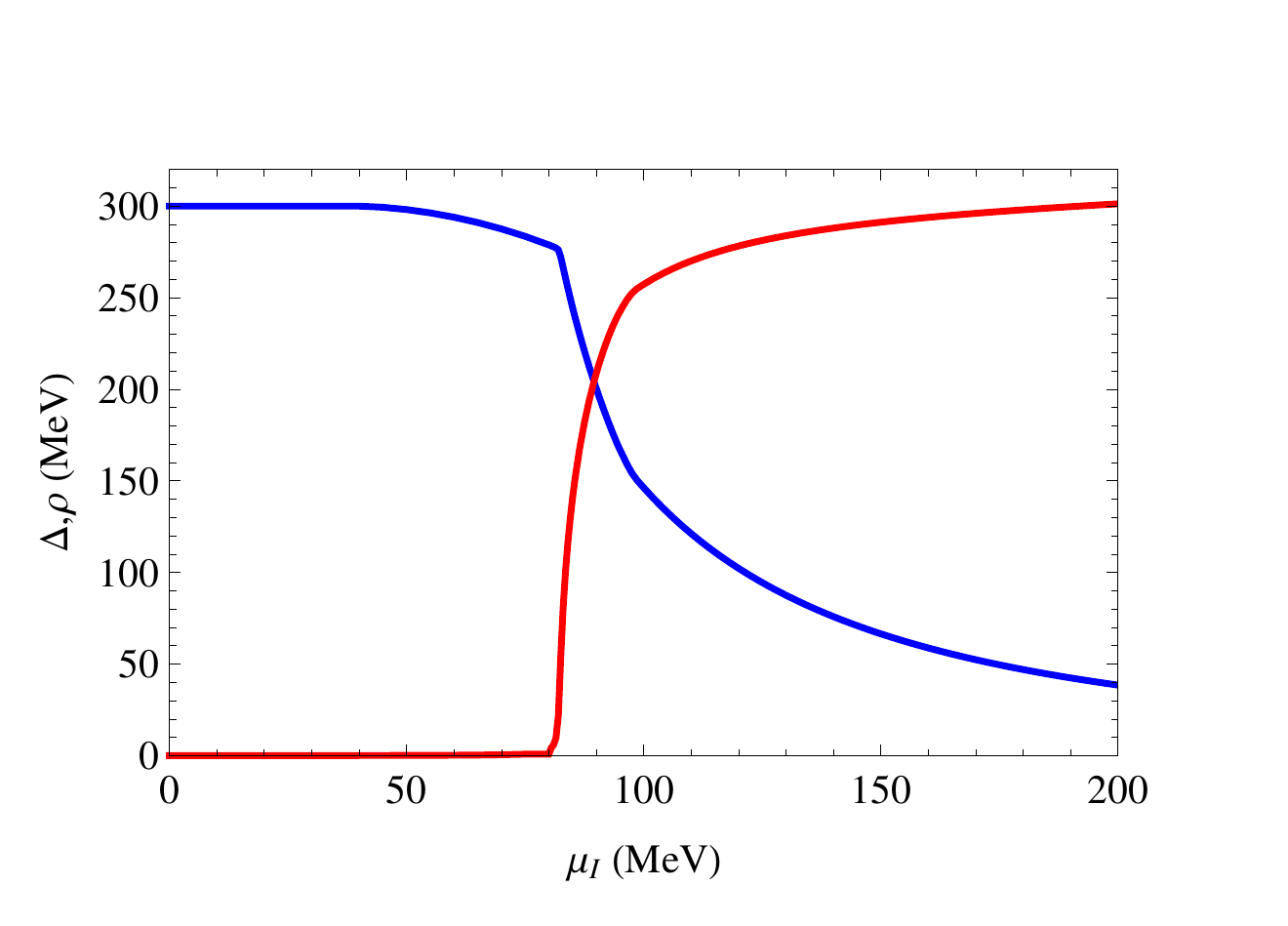}
\caption{Chiral condensate (blue line) and pion condensate (red line)
as functions of the isospin chemical potential $\mu_I$ for $\mu=0$
(upper panel) and $\mu=260$ MeV (lower panel)
at the physical point and $T=0$.}
\label{cross1}
\end{figure}

In Fig. \ref{cross1}, we show in the upper panel, the
chiral condensate (blue line) and pion condensate (red line)
as functions of the isospin chemical potential $\mu_I$ for $\mu=0$
at vanishing temperature. The onset of pion condensation
is at $\mu_I={1\over2}m_{\pi}$. The pion condensate increases, while the chiral 
condensate decreases.
One can view this as a rotation of the chiral condensate into a pion 
condensate as $\mu_I$ increases.
For $\mu_I<{1\over2}m_{\pi}$, the chiral condensate
is constant, which reflects the Silver Blaze property of the vacuum
phase: physical quantities are independent of the isospin
chemical potential for $\mu_I<\mu_I^c$, where the critical chemical potential
is $\mu_I^c={1\over2}m_{\pi}$ \cite{cohen}. 
In the lower panel of Fig. \ref{cross1}, we also show the 
chiral condensate (blue line) and pion condensate (red line)
as functions of the isospin chemical potential $\mu_I$, now for
$\mu=260$ MeV. We notice that there is a region of $\mu$, where $\Delta$
decreases before the onset of pion condensation.
This corresponds to a phase which is different from the vacuum phase since
the thermodynamic observables (here the chiral condensate) depend on 
the isospin chemical potential. 

\begin{figure}[htb]
\begin{center}
\includegraphics[width=0.4\textwidth]{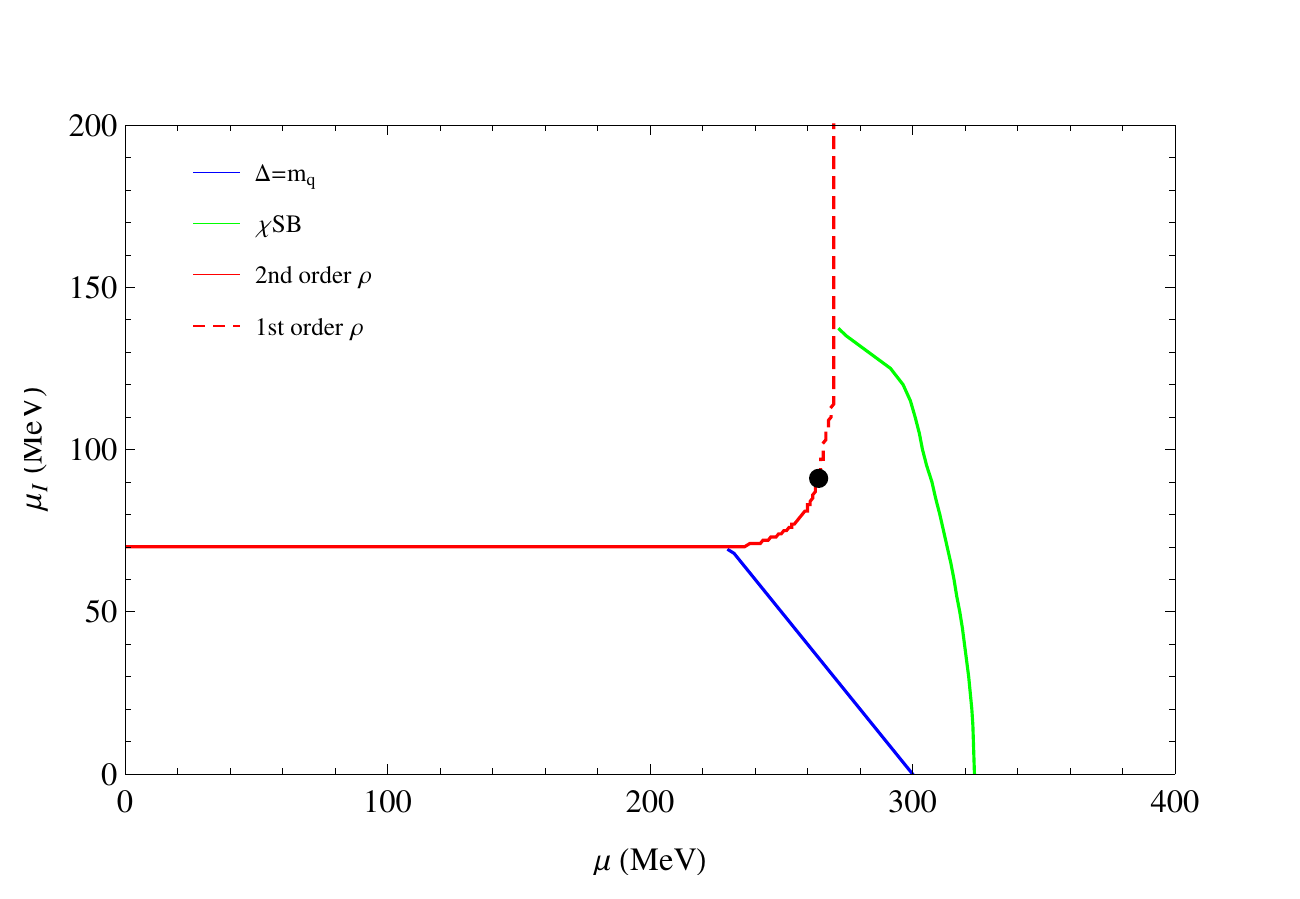}
\end{center}
\caption{Phase diagram 
in the $\mu$--$\mu_I$ plane at the physical point 
in the homogeneous case at $T=0$. See main text for details.}
\label{pionphase}
\end{figure}

More generally, the 
effective potential, and therefore thermodynamic observables are independent
of $\mu$ and $\mu_I$ in the region bounded by the 
$\mu_I$ and $\mu$ axes, and the straight lines given by 
$\mu+\mu_I=gf_{\pi}=m_q$ and $\mu_I=\mu_i^c={1\over2}m_{\pi}$.
In particular, the quark and isospin densities, which are defined by
\bqa
n_q=-{\partial V_{\rm full} \over\partial\mu}\;,
\hspace{1cm}
n_I=-{\partial V_{\rm full}\over\partial\mu_I}
\eqa 
vanish in this region. We therefore refer to this region as the vacuum phase.
This is shown in Fig. \ref{pionphase}, where the vacuum phase is 
bounded below
the solid blue and red lines. 
The red line shows the phase boundary between a phase with
$\rho=0$ and a pion-condensed phase. The transition is second order
when the red line is solid and first order when it is dashed. The
solid dot indicates the position of the critical end point where the
first-order line ends. The critical end point is located at
$(\mu, \mu_I) = (264,91)$ MeV. The green line indicates the boundary
between a chirally broken phase and a phase where chiral symmetry 
is approximately restored. 
For a sigma mass of $m_\sigma \geq 600 \text{ MeV}$ this transition is a 
crossover 
before it attaches to the dashed red 
line.\footnote{At the physical point, we define
the green line
by the inflection point of $\Delta$ as a function of $\mu$
for fixed $\mu_I$. 
For smaller sigma masses, the green line represents a 
first-order transition \cite{lorenz}. 
This line
ends at a critical point.} 
The region bounded by the three lines is a phase
with chiral symmetry breaking but no pion condensate. The 
effective potential depends on $\mu$ and $\mu_I$ and therefore
the quark and isospin densities are nonzero.

Since we have determined the parameters
of the Lagrangian such that the pion propagator including 
the self-energy has a pole at $m_{\pi}=140$ MeV, the onset of pion condensation
is exactly at $\mu_I={1\over2}m_{\pi}$ \cite{son}. 
We show this explicitly in Appendix C.
The result can be understood as follows. The 
energy of a zero-momentum pion in the vacuum phase is
$m_{\pi}-2\mu_I$. If it is a second-order transition it must take place
exactly at a point where the (medium-dependent) 
mass of the pion drops to zero because 
in the condensed phase there is a massless Goldstone mode associated with
the breaking of the $U(1)$ 
symmetry.\footnote{It is the $U_{I_3}(1)$
symmetry mentioned in Sec. II which is broken by the pion condensate.} 
If one uses matching at tree level, there will be finite corrections to this
relation. Likewise, if one uses the effective potential itself to define
the pion mass, one uses the pion self-energy at zero external momentum
and so the pole of the propagator is not at the physical mass. 
Again there will be finite corrections to $\mu_I={1\over2}m_{\pi}$ and
in some cases, the deviation can be substantial \cite{lorenz}.
Finally, we mention that the lattice result \cite{gergy1,gergy2,gergy3}
for the onset of 
pion condensation at $T=0$ for 2+1 dynamical quarks is in agreement with this.
Likewise, their simulations strongly suggest that the transition is second
order in the $O(2)$ universality class
in agreement with expectations.

\begin{figure}[htb]
\includegraphics[width=0.4\textwidth]{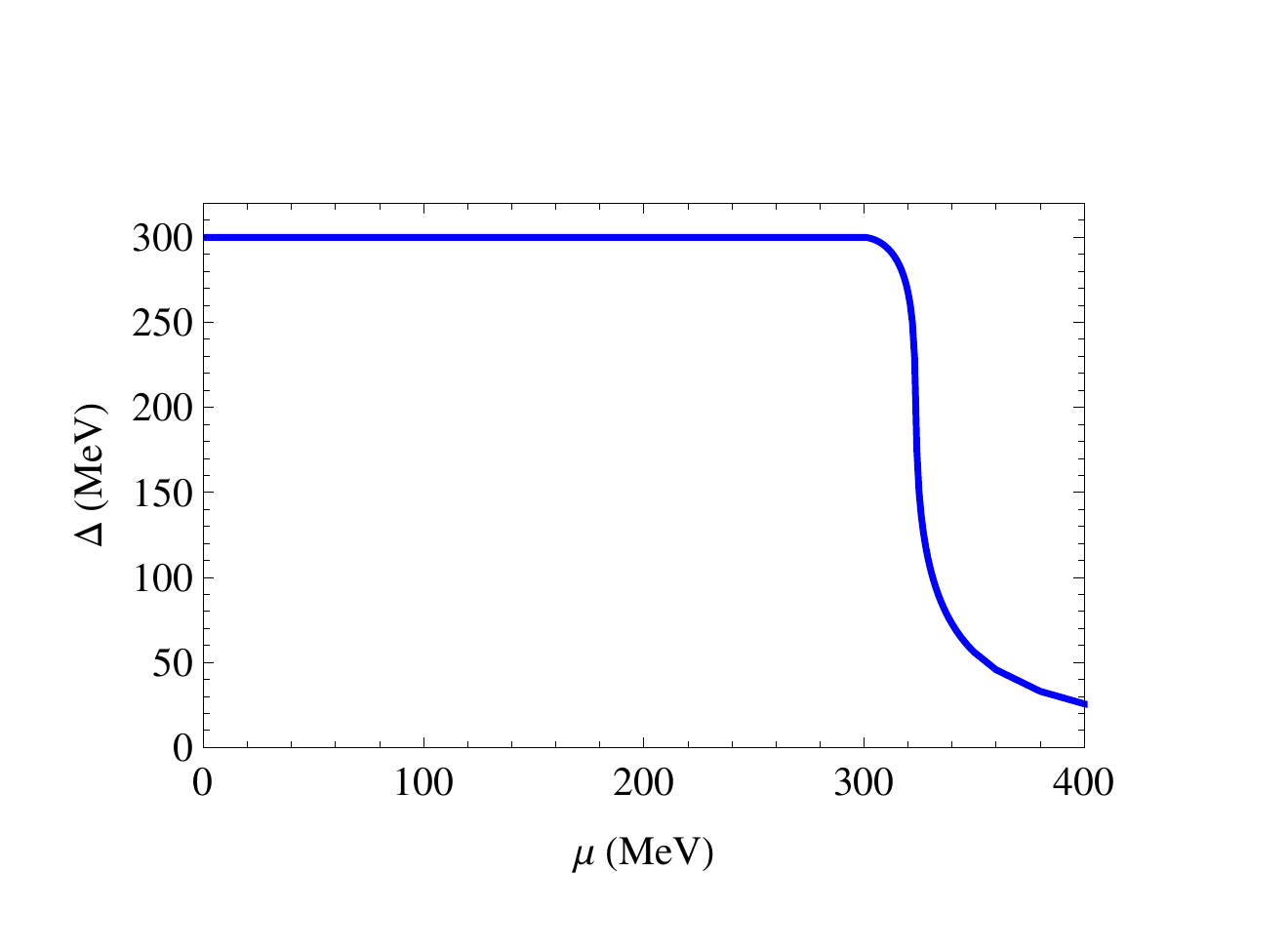}
\includegraphics[width=0.4\textwidth]{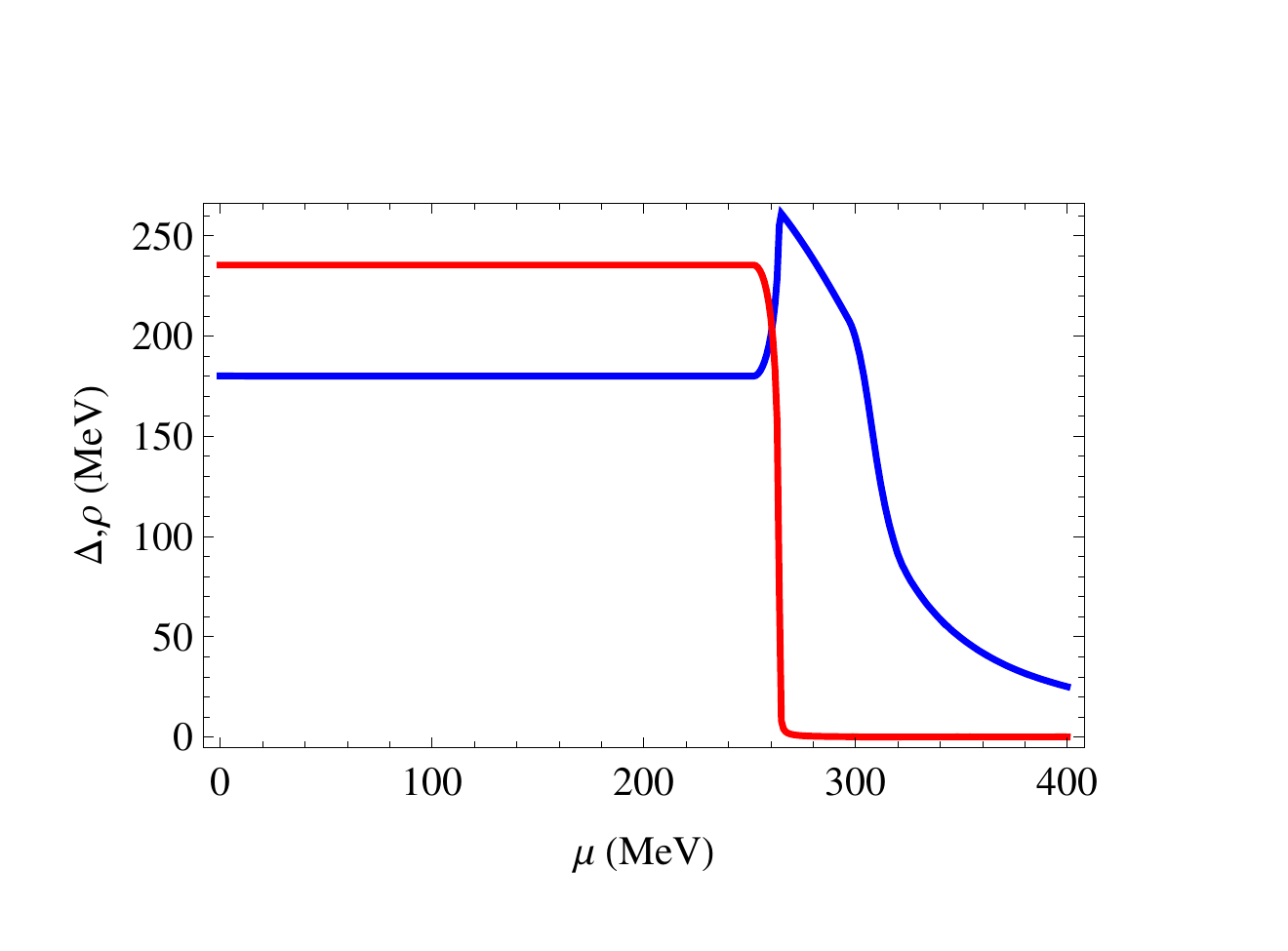}
\caption{Chiral condensate (blue line) and pion condensate (red line)
as functions of the chemical potential $\mu$ for $\mu_I=0$
(upper panel) and $\mu_I=90$ MeV (lower panel)
at the physical point and $T=0$.}
\label{cross2}
\end{figure}

In Fig. \ref{cross2}, we show $\Delta$ (blue line)
and $\rho$ (red line) as functions of the chemical potential for 
fixed value of the isospin chemical potential, $\mu_I=0$
in the upper panel and $\mu_I=90$ MeV in the lower panel. 
For $\mu_I=0$ there is no
pion condensate and there is a crossover at $\mu=323.6$ MeV.
For sigma masses below
600 MeV, this transition is first order.

\subsection{Inhomogeneous chiral condensate versus homogeneous pion condensate}
In this section, we generalize our result to nonzero $q$, i.e. we allow
for an inhomogeneous chiral condensate.

\begin{figure}[htb]
\includegraphics[width=0.4\textwidth]{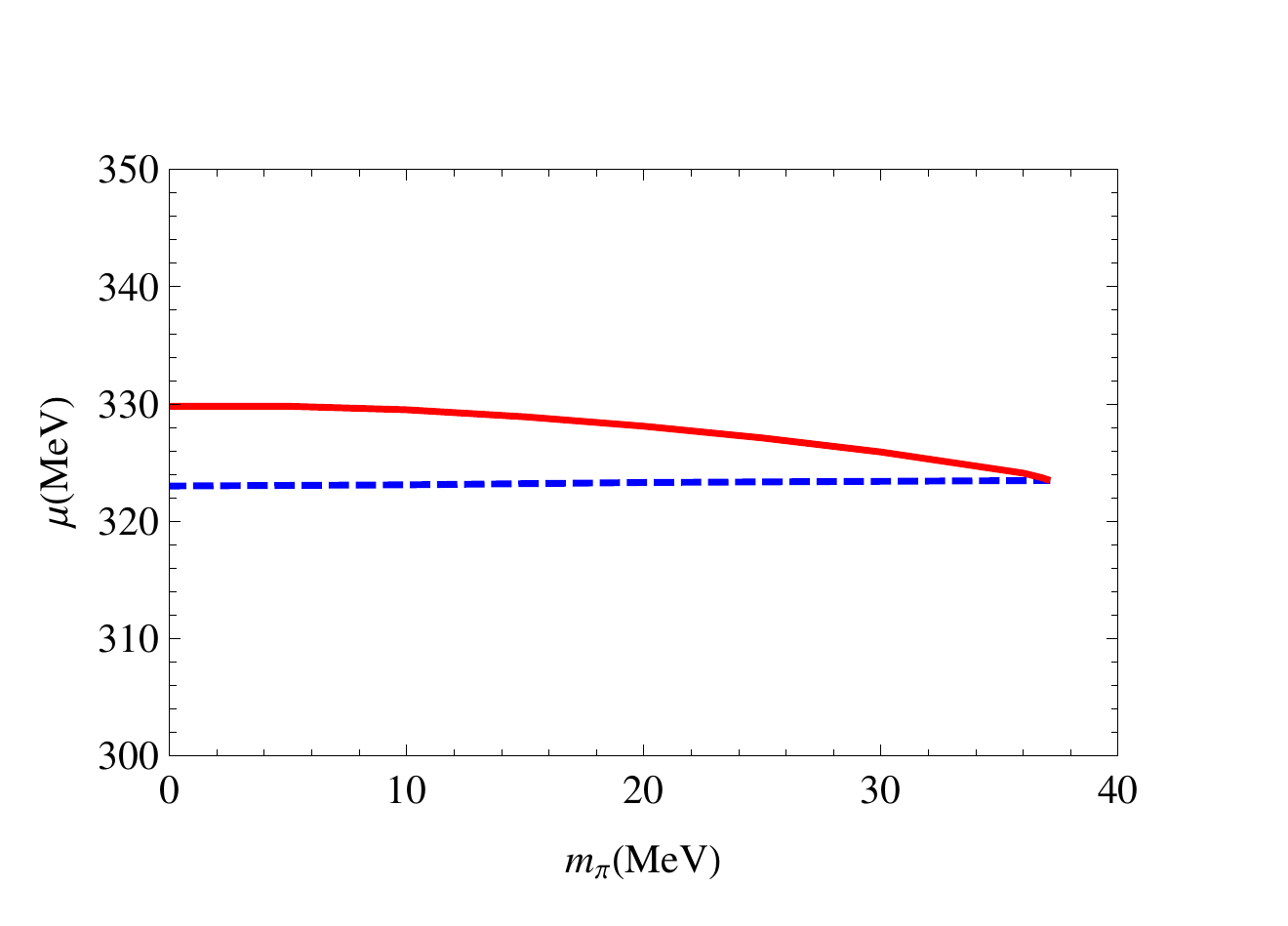}
\caption{Lower and upper limits of the chemical potential $\mu$
where an inhomogeneous phase exists as a function of $m_{\pi}$
for $\mu_I=0$.}
\label{boundary}
\end{figure}

It is known from earlier studies of inhomogeneous phases \cite{nick1} 
in the NJL and QM models that
the size of the region where an inhomogeneous phase exists depends
rather sensitively on the mass of the pion.
In Fig. \ref{boundary}, we show the lower and upper values of the
chemical potential $\mu$ for which an inhomogeneous phase
exists
as a function of $m_{\pi}$ at zero isospin chemical potential.
The curves meet at $m_{\pi}^c=37.1$ MeV, beyond which no inhomogeneous phase
exists. 
In particular, no inhomogeneous phase exists at the physical
point. 
The mechanism behind this is as follows. The 
symmetry-breaking  term $\sim -m_{\pi}^2f_{\pi}^2\cos(qz)\delta_{q,0}$ is nonzero
only for $q=0$. When the pion mass is sufficiently large, this term
is large enough for the homogeneous phase to be preferred over the 
inhomogeneous one.
This is in contrast to the results of \cite{nick1}, where
an inhomogeneous phase exists all the way up to $m_{\pi}=140$ MeV.
We can think of at least two reasons for this
qualtitative difference.
Firstly,  in Ref. \cite{nick1} tree-level parameters were used.
Secondly, a solitonic ansatz for the inhomogenity was
used and it is possible that this can sustain a inhomogeneous phase
for larger pion masses.

\begin{figure}[htb]
\includegraphics[width=0.4\textwidth]{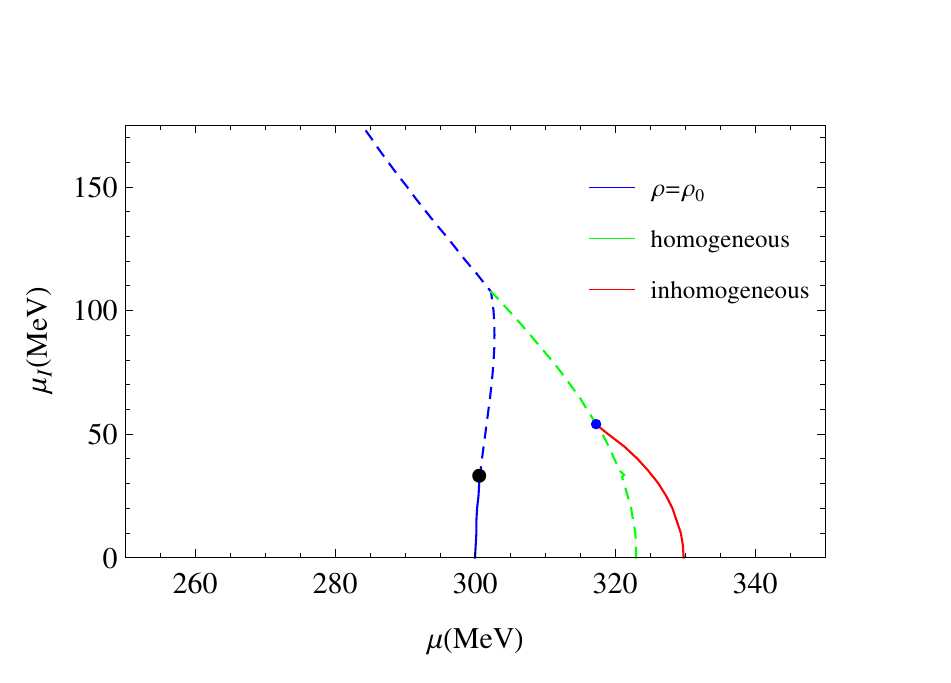}
\caption{Phase diagram in the $\mu$--$\mu_I$ plane at $T=0$ in the
chiral limit. See main text for details.}
\label{phase2}
\end{figure}

In Fig. \ref{phase2}, we show the phase diagram in the 
$\mu$--$\mu_I$ plane at vanishing temperature for $m_{\pi}=0$.
Solid lines indicate second-order transitions, while dashed lines indicate
first-order transitions. The black dot is the end point of 
the first-order line.
To the left of
the blue line there is a homogeneous 
pion condensate that does not change with increasing chemical 
potential and is equal to $\rho_0 = \rho(\mu=0,\mu_I)$. Between the blue and 
green line the pion condensate decreases and the quark density is non-zero. In 
both homogeneous phases we find $\Delta=0$, except for 
$\mu_I=0$.\footnote{On the $\mu$-axis, i.e. for $\mu_I=0$, the effective 
potential is a function of $\Delta^2 + \rho^2$ with a minimum at
$m_q^2$ up to $\mu=m_q$. 
Since the critical isospin chemical potential is $\mu_I^c=0$, 
pion condensation starts away from the $\mu$-axis and we chose $\Delta=m_q$ as 
the minimum.}
The region between the green and the red line is
the inhomogeneous phase, where
the chiral condensate and wave vector $q$ are nonzero. In this phase, the pion 
condensate
vanishes, implying that an inhomogeneous chiral condensate and
a homogeneous pion condensate do not coexist. 
Similar conclusions have been drawn in studies of the 1+1 dimensional
NJL model \cite{1p1klim,jens1+1}.
Finally, the region to the right of
red, blue, and green line segments is the symmetric phase, 
where $\Delta=\rho=q=0$.
The blue dot marks the Lifshitz point where the homogeneous, inhomogeneous and 
chirally symmetric phases connect. 

\begin{figure}[htb]
\includegraphics[width=0.4\textwidth]{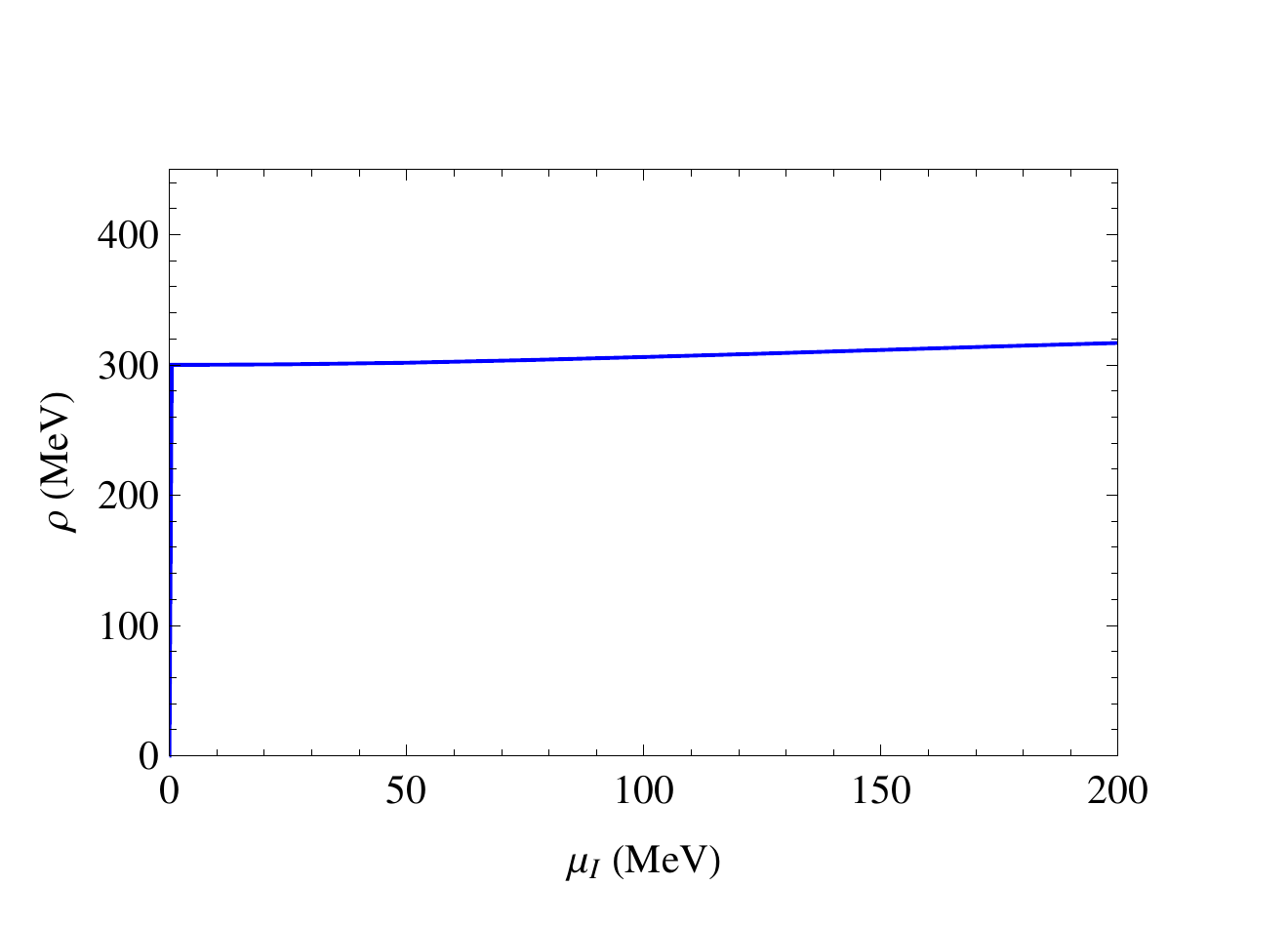}
\caption{Pion condensate as a function of the 
chemical potential $\mu_I$ in the chiral limit for $\mu=0$.}
\label{00}
\end{figure}

In Fig. \ref{00}, 
we show a cross section of the 
phase diagram in Fig. \ref{phase2} 
in the chiral limit
for $\mu=0$. The pion condensate is shown as a function  of 
the isospin
chemical potential $\mu_I$. We notice that the chiral condesate
in the vacuum immediately vanishes once $\mu_I>0$ and is rotated to
a pion condensate with the value $\rho=m_q$. The pion condensate
increases further as we increase the isospin chemical potential $\mu_I$.

\begin{figure}[htb]
\includegraphics[width=0.4\textwidth]{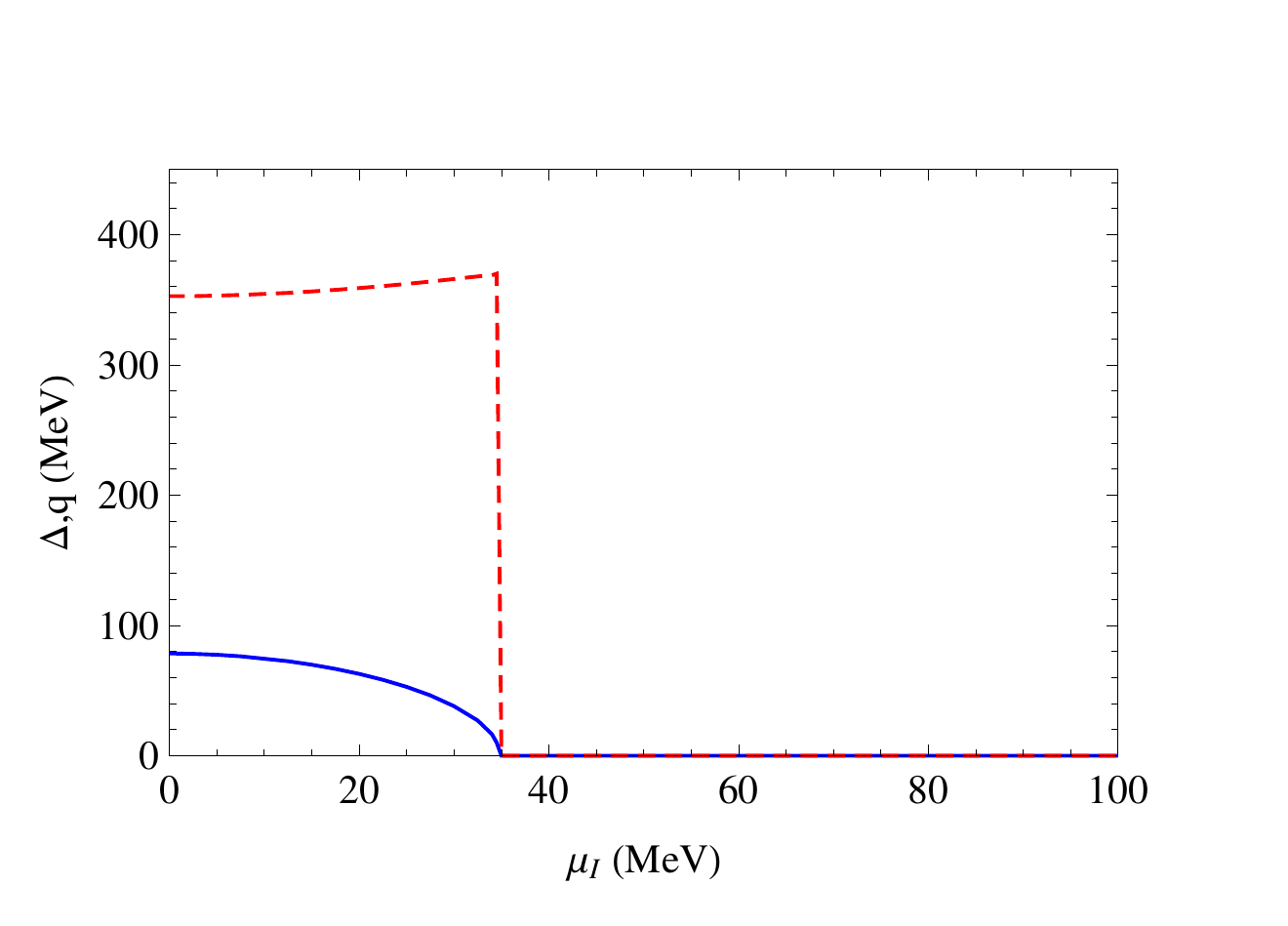}
\caption{Chiral condensate $\Delta$
(blue line) and wave vector $q$ (red line) as a function of the isospin
chemical potential $\mu_I$ in the chiral limit for $\mu=325$ MeV.}
\label{325}
\end{figure}

In Fig. \ref{325}, we show a cross section of the phase diagram 
in Fig. \ref{phase2} in the chiral limit
for $\mu=325$ MeV. The chiral condensate
is the blue line and the wave vector is the red line. The
transition to the symmetric phase is of second order.
Note that the pion condensate is zero in the inhomogeneous phase.

\begin{figure}[htb]
\includegraphics[width=0.4\textwidth]{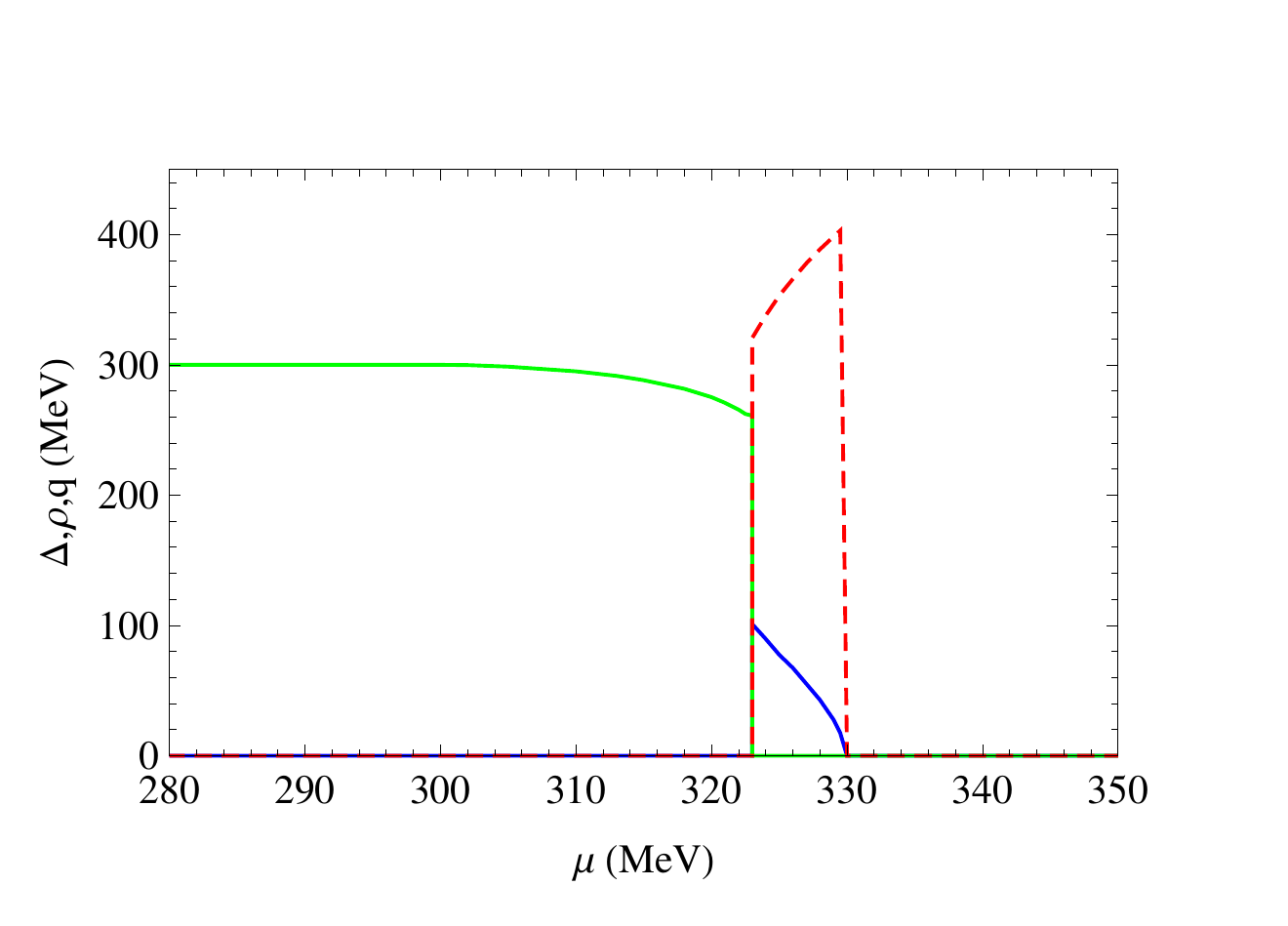}
\caption{Chiral condensate (blue line), wave vector (red line), and
pion condensate (green line) as functions of the 
chemical potential $\mu$ in the chiral limit for $\mu_I=5$ MeV.}
\label{10}
\end{figure}

In Fig. \ref{10}, we show the 
chiral condensate (blue line), wave vector (red line), and
pion condensate (green line) as functions of the 
chemical potential $\mu$ in the chiral limit for $\mu_I=5$ MeV.
In the homogeneous phase, a pion condensate is favored over a chiral condensate.
At the first-order transition at $\mu=323$ MeV, the pion condensate drops to
zero and we enter a phase with an inhomogeneous chiral condensate.
At $\mu=329.8$ MeV, there is a second-order transition to the symmetric phase.

\begin{figure}[htb]
\includegraphics[width=0.4\textwidth]{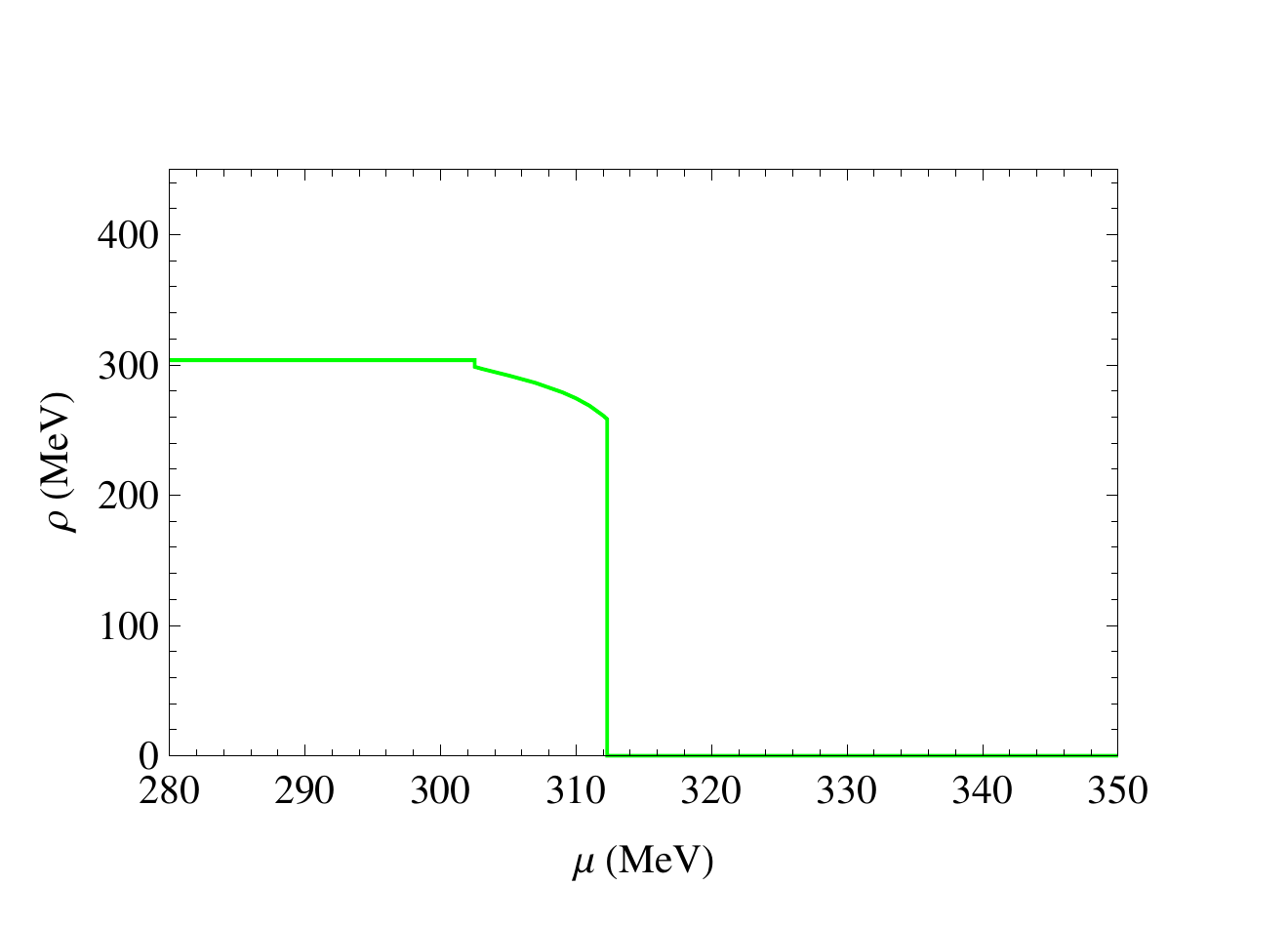}
\caption{Pion condensate $\rho$ as a function of the 
chemical potential $\mu$ in the chiral limit for $\mu_I=75$ MeV.}
\label{150}
\end{figure}

In Fig. \ref{150}, we show the pion condensate as a function of
the chemical potential in the chiral limit and for $\mu_I=75$ MeV.
There are two first-order transitions at $\mu=302.5$ MeV and 
$\mu=312.3$ MeV, respectively,
where the value of the pion condensate jumps discontinously.

\section{Summary}
In the present paper, we have studied the phases of QCD at $T=0$
in the $\mu$--$\mu_I$ plane using the quark-meson model as a low-energy
effective model. Combining
the $\overline{\rm MS}$ and ${\rm OS}$ schemes, we have
determined the parameters of the model, whose values are consistent
with the approximation that we used for the effective potential.
In contrast to other model calculations, where the parameters are fixed
at tree level, our method guarantees that the critical isospin
chemical potential at $T=0$ is exactly at $\mu_I={1\over2}m_{\pi}$.

Moreover, we found that the existence of an inhomogeneous chiral condensate
depends on the value of the pion mass, which is in agreement with
earlier model calculations \cite{nick1}. Specifically, we found that the
chiral density wave is disfavored for pion masses larger than 
approximately
$37$ MeV. The existence of such a critical pion mass is in contrast
to the results of Ref.~\cite{nick1}, where a inhomogeneous chiral
condensate exists all the way up to $m_{\pi}=140$ MeV.
The difference is probably due to 
tree-level versus one-loop matching of the parameters
as well as different ans\"atze
for the inhomogeneity.

Finally, we mapped out the phase diagram at $T=0$ in the $\mu$--$\mu_I$
plane both in the chiral limit and at the physical point.
In the chiral limit, we have seen that there is a region 
where an inhomogeneous chiral condensate
is favored. The corresponding Lifshitz point is a tricritical point,
as three phases meet.
At the physical point, the phase diagram is in qualitative agreement with the
mean-field calculations in \cite{lorenz}. However, they have also
performed a functional renormalization group (FRG)
calculation of the phase diagram
and including the mesonic fluctuations gives rise to 
qualitative differences. One important
difference is that the transition to a
Bose-condensed phase is second order in the entire $\mu$--$\mu_I$ plane.
It would be very interesting to perform an FRG calculation including
inhomogeneous phases to see whether the mean-field picture 
found here will change.


\section*{Acknowledgments}
The authors would like to thank Tomas Brauner for useful discussions.
P. K. would like to thank the Department of Physics at NTNU for
kind hospitality during the latter stages of this work.

\appendix
\section{Integrals}
With dimensional regularization,
the momentum integral is generalized to $d=3-2\epsilon$
spatial dimensions. We define the dimensionally regularized integral by
\bqa
\int_p&=&
\left({e^{\gamma_E}\Lambda^2\over4\pi}\right)^{\epsilon}
\int{d^{d}p\over(2\pi)^d}\;,
\label{dint}
\eqa
where $\Lambda$ is the renormalization scale in the
modified minimal subtraction scheme $\overline{\rm MS}$.
It is convenient to write the integral as 
\bqa
\int_p&=&
\left({e^{\gamma_E}\Lambda^2\over4\pi}\right)^{\epsilon}
\int{d^{d-1}p_{\perp}\over(2\pi)^{d-1}}
\int_{-\infty}^{\infty}{dp_{\parallel}\over2\pi}\;.
\label{dint2}
\eqa
\begin{widetext}
In order to calculate the effective potential, we need the vacuum integrals
\bqa
\int_p\sqrt{p^2+M^2}
&=&-{M^4\over(4\pi)^2
}\left({e^{\gamma_E}\Lambda^2\over M^2}\right)^{\epsilon}
\Gamma(-2+\epsilon)
=-{M^4\over2(4\pi)^2
}\left({\Lambda^2\over M^2}\right)^{\epsilon}
\left[{1\over\epsilon}+{3\over2}+{\cal O}(\epsilon)\right]\;,
\label{i1}
\\ 
\int_p{1\over\sqrt{p^2+M^2}}
&=&-{2M^2\over(4\pi)^2}
\left({e^{\gamma_E}\Lambda^2\over M^2}\right)^{\epsilon}
\Gamma(-1+\epsilon)=-{2M^2\over(4\pi)^2}
\left({\Lambda^2\over M^2}\right)^{\epsilon}
\left[{1\over\epsilon}+1+{\cal O}(\epsilon)\right]\;,
\label{i2}
\\ 
\int_p{1\over(p^2+M^2)^{3\over2}}
&=&{4\over(4\pi)^2}
\left({e^{\gamma_E}\Lambda^2\over M^2}\right)^{\epsilon}
\Gamma(\epsilon)
={4\over(4\pi)^2}
\left({\Lambda^2\over M^2}\right)^{\epsilon}
\left[{1\over\epsilon}+{\cal O}(\epsilon)\right]\;,
\label{i3}
\\
\int_p{(p_{\perp}^2+\rho^2)\over(p^2+\Delta^2+\rho^2)^{3\over2}}
&=&-{4\Delta^2\over(4\pi)^2}
\left({e^{\gamma_E}\Lambda^2\over\Delta^2+\rho^2}\right)^{\epsilon}
\Gamma(\epsilon)
=-{4\Delta^2\over(4\pi)^2}
\left({\Lambda^2\over\Delta^2+\rho^2}\right)^{\epsilon}
\left[{1\over\epsilon}+{\cal O}(\epsilon)\right]\;,
\\ \nonumber
\int_p{(p_{\perp}^2+\rho^2)(4\Delta^2-\rho^2+4p_{\parallel}^2-p_{\perp}^2)
\over(p_{\perp}^2+p_{\parallel}^2+M^2)^{7\over2}}
&=&{16\over3(4\pi)^2}
\left({e^{\gamma_E}\Lambda^2\over\Delta^2+\rho^2}\right)^{\epsilon}
{\Delta^2\over(\Delta^2+\rho^2)^2}
\left[(1-\epsilon)\Delta^2+2\rho^2\right]\Gamma(1+\epsilon)
\\ &=&
{16\over3(4\pi)^2}{\Delta^2(\Delta^2+2\rho^2)\over(\Delta^2+\rho^2)^2}
\left[1+{\cal O}(\epsilon)\right]
\;,
\\ \nonumber
\int_p{(4\Delta^2-\rho^2+4p_{\parallel}^2-p_{\perp}^2)
\over(p^2+\Delta^2+\rho^2)^{7\over2}}
&=&{16\over3(4\pi)^2}
\left({e^{\gamma_E}\Lambda^2\over\Delta^2+\rho^2}\right)^{\epsilon}
{\Delta^2\over(\Delta^2+\rho^2)^2}
\Gamma(2+\epsilon)
\\ 
&=&{16\over3(4\pi)^2}
{\Delta^2\over(\Delta^2+\rho^2)^2}\left[1+{\cal O}(\epsilon)\right]
\;.
\eqa
\section{Parameter fixing}
In this Appendix, we briefly discuss the fixing of the model parameters.
At tree level, the relations between these parameters and the
physical quantities are given by Eqs. (\ref{rel1})--(\ref{rel2}).
In the
on-shell scheme, the divergent loop integrals are regularized 
using dimensional regularization, but the counterterms are defined
differently from those in the minimal subtraction
scheme.   The  counterterms  in  the  on-shell  scheme  are
chosen so that they exactly cancel the loop corrections
to the self-energies \footnote{
And such that the residues of the propagators evaluated on the mass shell
are unity.}
and couplings evaluated on the mass shell, and
consequently, the renormalized parameters are independent
of the renormalization scale and satisfy the tree-level 
relations \cite{sir1,sir2,hollik}.
In the $\overline{\rm MS}$ scheme, the counterterms are chosen
so that they cancel only the poles in $\epsilon$ of the loop corrections.
The bare parameters
are the same in the two schemes and so we can relate
the corresponding renormalized parameters.
The running parameters in the $\overline{\rm MS}$ scheme
can therefore be expressed in terms of the 
physical masses $m_{\sigma}$, $m_{\pi}$, and $m_q$ as well the pion decay constant.
In Ref. \cite{crew} we found 
\bqa \nonumber
m^2_{\ms} 
&=& m^2 +8ig^2N_c \left[ A(m_q^2) +\mbox{$1\over4$}(m_\sigma^2-4m_q^2)
B(m_\sigma^2) 
-\mbox{$3\over4$}m_{\pi}^2B(m_{\pi}^2)\right] -\delta m^2_{\ms} \\
 &=& m^2+
\dfrac{4g^2N_c}{(4\pi)^2} 
\left[m^2\log\mbox{$\Lambda^2\over m_q^2$}
-2m_q^2-{1\over2}\left(m_\sigma^2-4m_q^2\right)F(m_\sigma^2) 
+{3\over2}{m_\pi^2}F(m_\pi^2) \right]\;, 
\label{osm1}
\\ \nonumber
\lambda_{\ms} 
&=& \lambda -\dfrac{12ig^2N_c}{f_{\pi}^2} (m_\sigma^2-4m_q^2)B(m_\sigma^2) 
+\dfrac{12ig^2N_c}{f_{\pi}^2}m_\pi^2B(m_\pi^2) 
-4i\lambda g^2N_c\left[B(m_{\pi}^2)+m_{\pi}^2B^{\prime}(m_{\pi}^2)\right]
- \delta\lambda_{\ms}
\\ \nonumber &=& 
\lambda
+\bigg\{\dfrac{12g^2N_c}{(4\pi)^2f_\pi^2}\left[(m_\sigma^2-4m_q^2)\left(
\log\mbox{$\Lambda^2\over m_q^2$} +F(m_\sigma^2)\right)
+m_\sigma^2\left(
\log\mbox{$\Lambda^2\over m_q^2$}+F(m_\pi^2)+m_\pi^2F^{\prime}(m_\pi^2)\right)
\right.\\ &&\left.
-m_\pi^2\left(
2\log\mbox{$\Lambda^2\over m_q^2$}+2F(m_\pi^2)
+F^{\prime}(m_\pi^2)\right)\right]\bigg\}\;,
\label{osl}
\\ 
g_{\ms}^2  &= &g^2-4ig^4N_c
\left[B(m_\pi^2) +m_\pi^2B^\prime(m_\pi^2) \right] -\delta g_{\ms}^2 
={m_q^2\over f_{\pi}^2}\left\{1 + \dfrac{4g^2N_c}{(4\pi)^2}\left[
\log\mbox{$\Lambda^2\over m_q^2$}+F(m_\pi^2) +m_\pi^2F^\prime(m_\pi^2)
\right]\right\}\;,
\label{g00}
\\ 
h_{\ms}&=&h-2ig^2N_cm_\pi^2f_\pi\left[
B(m_\pi^2)-m_\pi^2B^{\prime}(m_\pi^2)\right]-\delta h_{\ms}
=h
\left\{
1+{2g^2N_c\over(4\pi)^2}
\left[\log\mbox{$\Lambda^2\over m_q^2$}
+F(m_\pi^2)-m_\pi^2F^{\prime}(m_\pi^2)\right]\right\}\;,
\label{hhh}
\eqa
\end{widetext}
where $A(m_q^2)$, $B(p^2)$, and $B^{\prime}(p^2)$
are integrals in $d=4-2\epsilon$ dimensions in Minkowski space. 
Going to Euclidean space, they can be straightforwardly computed and
read
\bqa\nonumber
A(m^2_q)&=&\int_p{1\over p^2-m^2_q}
\\&=&{im^2_q\over(4\pi)^2}\left({\Lambda^2\over m_q^2}\right)
\left[{1\over\epsilon}+1+{\cal O}(\epsilon)\right]\;,
\label{adef}
\\ \nonumber
B(p^2)&=&
\int_p{1\over(k^2-m_q^2)[(k+p)^2-m_q^2]}
\\ &=&
{i\over(4\pi)^2}\left({\Lambda^2\over m_q^2}\right)
\left[{1\over\epsilon}+F(p^2)
+{\cal O}(\epsilon)\right]\;,
\label{bdef}
\\
B^{\prime}(p^2)&=&{i\over(4\pi)^2}F^{\prime}(p^2)\;.
\eqa
Here we have defined 
\bqa
F(p^2)&=&2-2r\arctan\left({1\over r}\right)\;,
\\
F^{\prime}(p^2)&=&{4m_q^2r\over p^2(4m_q^2-r^2)}\arctan\left({1\over r}\right)
-{1\over p^2}\;,
\eqa
with $r=\sqrt{{4m_q^2\over p^2}-1}$.


The running parameters satisfy 
the following renormalization group equations
\bqa
\label{run1}
\Lambda{dm_{\ms}^2(\Lambda)\over d\Lambda}&=&{8 N_c m^2_{\ms}(\Lambda)g_{\ms}^2
(\Lambda)
\over(4\pi)^2}\;,
\\
\Lambda{dg_{\ms}^2(\Lambda)\over d\Lambda}&=&{8 N_c g_{\ms}^4(\Lambda)
\over(4\pi)^2}
\;,
\\
\Lambda{d\lambda_{\ms}(\Lambda)\over d\Lambda}&=&{16N_c\over(4\pi)^2}
\left[\lambda_{\ms}(\Lambda) g^2_{\ms}(\Lambda)-6g^4_{\ms}(\Lambda)
\right]\;,
\\
\Lambda{dh_{\ms}(\Lambda)\over d\Lambda}&=&{4 N_c g^2_{\ms}(\Lambda)
h_{\ms}(\Lambda)\over(4\pi)^2}
\label{run4}
\;.
\eqa
The solutions to Eqs. (\ref{run1})--(\ref{run4}) are
\bqa
\label{sol1}
m_{\ms}^2(\Lambda)&=&
{m_0^2\over1-{4g_0^2N_c\over(4\pi)^2}
\log{\Lambda^2\over \Lambda_0^2}
}\;.
\\
g_{\ms}^2(\Lambda)&=&
{g_0^2\over1-{4g_0^2N_c\over(4\pi)^2}
\log{\Lambda^2\over \Lambda_0^2}
}\;,
\\
\lambda_{\ms}(\Lambda)&=&{\lambda_0-{48g_0^4N_c\over(4\pi)^2}
\log{\Lambda^2\over \Lambda_0^2}
\over\left(1-{4g_0^2N_c\over(4\pi)^2}
\log{\Lambda^2\over \Lambda_0^2}
\right)^2}\;,
\label{sol3}
\\
h_{\ms}(\Lambda)&=&
{h_0\over1-{2g_0^2N_c\over(4\pi)^2}
\log{\Lambda^2\over \Lambda_0^2}
}\;,
\label{sol5}
\eqa
where $m_0^2$, $g_0^2$, $\lambda_0$ and $h_0$, are the 
values of the running parameters at the scale $\Lambda_0$.
We choose $\Lambda_0$ to satisfy
\bqa
\log{\Lambda_0^2\over m_q^2}+F(m_\pi^2)+m_\pi^2F^{\prime}(m_\pi^2)
&=&0\;.
\label{l0}
\eqa
$F(m_\pi^2)$ and $m_\pi^2F^{\prime}(m_\pi^2)$ vanish in the chiral limit which
implies that $\Lambda_0=m_q$.
We can now evaluate Eqs. (\ref{osm1})--(\ref{hhh}) at $\Lambda=\Lambda_0$ 
to find
$m_0^2$, $\lambda_0$, $g_0^2$, and $h_0$. Inserting Eqs. 
(\ref{sol1})--(\ref{sol5}) into Eq. (\ref{veff})
using the results for $m_0^2$, $\lambda_0$, $g_0^2$, and $h_0$, 
we obtain the final result Eq. (\ref{fullb}).
\section{Onset of pion condensation}
In order to show that the onset of pion condensation is exactly at
$\mu_I={1\over2}m_{\pi}$, 
we expand the one-loop contribution to the free energy in powers
of $\rho$ to fourth order to obtain an effective Ginzburg-Landau energy
functional. For simplicity, we consider the case $\mu=0$ where
the finite-density contribution (\ref{fd}) vanishes.

After renormalization, the one-loop effective potential will be
of the form
\bqa
V_{\rm 1-loop}&=&\alpha_0+\alpha_2\rho^2+\alpha_4\rho^4\;,
\eqa
where the coefficients $\alpha_i$ depend on the physical
quantities $m_{\sigma}$, $m_{\pi}$, $f_{\pi}$, $\mu_I$ and $\Delta$.
The critical isospin chemical potential $\mu_I^c$
is defined by
$\alpha_2=0$ evaluated at $\Delta=m_q$, i.e. its vacuum value.
If $\alpha_4>0$ at $\mu_I^c$ and $\Delta=m_q$, then there is a second
order transition at $\mu_I^c$.

\begin{widetext}
Setting $q=0$ in Eq. (\ref{v1}) the one-loop 
vacuum energy reduces to 
\bqa
V_{1}&=&-2N_c\int_p
\sqrt{\left(\sqrt{p^2+\Delta^2}\pm{\mu_I}\right)^2+\rho^2}\;.
\label{exp}
\eqa
Expanding Eq. (\ref{exp}) 
in powers of
$\rho$ up to fourth order, we find 
\bqa
\label{rexp}
V_{1}&=&-2N_c\int_p\left[\sqrt{p^2+\Delta^2}\pm{\mu_I}
+{1\over2}{\rho^2\over\sqrt{p^2+\Delta^2}\pm{\mu_I}}
-{1\over8}{\rho^4\over(\sqrt{p^2+\Delta^2}\pm{\mu_I})^{3}}\right]\;.
\eqa
In order to isolate the divergence, we expand the different terms
in Eq. (\ref{rexp}) in powers
of $\mu_I$ to the appropriate power order, then 
add and subtract suitable terms. This yields
\bqa
V_1&=&-4N_c\int_p\left[\sqrt{p^2+\Delta^2}
+{1\over2}\rho^2\left({1\over\sqrt{p^2+\Delta^2}}
+{\mu_I^2\over(p^2+\Delta^2)^{3\over2}}\right)
-{1\over8}{\rho^4\over\left(p^2+\Delta^2\right)^{3\over2}}
\right]+V_{\rm fin,1}+V_{\rm fin,2}\;,
\eqa
where the finite terms can be calculated directly in $d=3$ and read
\bqa\nonumber
V_{\rm fin,1}&=&-2N_c\rho^2\int_p\left[
{\sqrt{p^2+\Delta^2}\over p^2+\Delta^2-{\mu_I}^2}
-{1\over\sqrt{p^2+\Delta^2}}
-{\mu_I^2\over(p^2+\Delta^2)^{3\over2}}
\right] \\
&=&
-{16N_c\over(4\pi)^2}\rho^2\mu_I^2\left[
1-s\arctan\left({1\over s}\right)
\right]\;,\\ \nonumber
V_{\rm fin,2}&=&{1\over4}N_c\rho^4\int_p\left[
{1\over\left(\sqrt{p^2+\Delta^2}\pm\mu_I\right)^3}
-{2\over\left(p^2+\Delta^2\right)^{3\over2}}
\right] \\
&=&
{2N_c\over(4\pi)^2}\rho^4
\left[
{1\over s^2}+{1\over s^3}\left(
{3\Delta^2\over\mu_I^2}-2\right)
\arctan\left({1\over s}\right)
\right]\;,
\eqa
where $s=\sqrt{{\Delta^2\over\mu_I^2}-1}$\;.
In analogy with Eq. (\ref{veff}), we obtain the renormalized vacuum
potential through order $\rho^4$ 
\bqa\nonumber
V_{\rm 1-loop}&=&
{1\over2}{m^2_{\ms}\over g^2_{\ms}}\Delta^2
+{1\over2}{m^2_{\ms}-4\mu_I^2\over g^2_{\ms}}\rho^2
+{\lambda_{\ms}\over24g_{\ms}^4}\left(\Delta^2+\rho^2\right)^2
-{h_{\ms}\over g_{\ms}}\Delta\cos(qz)\delta_{q,0}\;.
\\ 
&&+{2N_c\over(4\pi)^2}
\left\{(\Delta^2+\rho^2)^2\log{\Lambda^2\over\Delta^2}
+4\rho^2\mu_I^2\log{\Lambda^2\over\Delta^2}
+{3\over2}\Delta^4+2\Delta^2\rho^2
\right\}
+V_{\rm fin,1}+V_{\rm fin,2}\;.
\eqa
Using Eqs. (\ref{i1})--(\ref{i3}) as well as 
(\ref{sol1})--(\ref{sol5}), the
renormalized effective potential up to order $\rho^4$ is
\bqa\nonumber
V_{\rm 1-loop}&=&
\dfrac{3}{4}m_\pi^2 f_\pi^2
\left\{1-\dfrac{4 m_q^2N_c}{(4\pi)^2f_\pi^2}m_\pi^2F^{\prime}(m_\pi^2)
\right\}\dfrac{\Delta^2+\rho^2}{m_q^2}
\\ \nonumber &&
 -\dfrac{1}{4}m_\sigma^2 f_\pi^2
\left\{
1 +\dfrac{4 m_q^2N_c}{(4\pi)^2f_\pi^2}
\left[ \left(1-\mbox{$4m_q^2\over m_\sigma^2$}
\right)F(m_\sigma^2)
 +\dfrac{4m_q^2}{m_\sigma^2}
-F(m_\pi^2)-m_\pi^2F^{\prime}(m_\pi^2)
\right]\right\}\dfrac{\Delta^2+\rho^2}{m_q^2} 
\\ && \nonumber
-2\mu_I^2f_{\pi}^2
\left\{
1-{4m_q^2N_c\over(4\pi)^2f_{\pi}^2}
\left[
\log{\Delta^2\over m_q^2}
+F(m_{\pi}^2)
+m_{\pi}^2F^{\prime}(m_{\pi}^2)
-2+2s\arctan{1\over s}
\right]
\right\}{\rho^2\over m_q^2}
\\  \nonumber
 & & + \dfrac{1}{8}m_\sigma^2 f_\pi^2
\left\{ 1 -\dfrac{4 m_q^2  N_c}{(4\pi)^2f_\pi^2}\left[
\dfrac{4m_q^2}{m_\sigma^2}
\log\mbox{$\Delta^2\over m_q^2$}
-\left( 1 -\mbox{$4m_q^2\over m_\sigma^2$}\right)F(m_\sigma^2)
+F(m_\pi^2)+m_\pi^2F^{\prime}(m_\pi^2)\right]
 \right\}\dfrac{(\Delta^2+\rho^2)^2}{m_q^4}
\\ \nonumber&&
- \dfrac{1}{8}m_\pi^2 f_\pi^2
\left[1-\dfrac{4 m_q^2N_c}{(4\pi)^2f_\pi^2}m_\pi^2F^{\prime}(m_\pi^2)\right]
\dfrac{(\Delta^2+\rho^2)^2}{m_q^4}
-m_\pi^2f_\pi^2\left[
1-\dfrac{4 m_q^2  N_c}{(4\pi)^2f_\pi^2}m_\pi^2F^{\prime}(m_\pi^2)
\right]\dfrac{\Delta}{m_q}
\\
&&+{2N_c\over(4\pi)^2}
\left\{
{3\over2}\Delta^4+2\Delta^2\rho^2
+\left[
{1\over s^2}+{1\over s^3}\left(
{3\Delta^2\over\mu_I^2}-2\right)
\arctan\left({1\over s}\right)\right]\rho^4
\right\}
\;.
\label{rhoex}
\eqa
From the effective potential Eq. (\ref{rhoex}), we can 
read off the coefficients $\alpha_2$ and $\alpha_4$. Evaluated
at $\Delta=m_q$, we find
\bqa
\alpha_2&=&
\dfrac{1}{2}
{f_\pi^2\over m_q^2}
\left\{
m_\pi^2\left[
1-\dfrac{4 m_q^2N_c}{(4\pi)^2f_\pi^2}m_\pi^2F^{\prime}(m_\pi^2)\right]
-4\mu_I^2\left[
1-{4m_q^2N_c\over(4\pi)^2f_{\pi}^2}
\bigg(
F(m_{\pi}^2)
+m_{\pi}^2F^{\prime}(m_{\pi}^2)-F(4\mu_I^2)\bigg)
\right]
\right\}\;,
\\ \nonumber
\alpha_4&=&
\dfrac{1}{8}{m_\sigma^2 f_\pi^2\over m_q^4}
\left\{ 1 -\dfrac{4 m_q^2  N_c}{(4\pi)^2f_\pi^2}\left[
-\left( 1 -\mbox{$4m_q^2\over m_\sigma^2$}\right)F(m_\sigma^2)
+F(m_\pi^2)+m_\pi^2F^{\prime}(m_\pi^2)\right]
 \right\}- \dfrac{1}{8}{m_\pi^2 f_\pi^2\over m_q^4}
\left[1-\dfrac{4 m_q^2N_c}{(4\pi)^2f_\pi^2}m_\pi^2F^{\prime}(m_\pi^2)\right]
\\ 
&&
+{2N_c\over(4\pi)^2}
\left[
{1\over s^2}+{1\over s^3}\left(
{3\Delta^2\over\mu_I^2}-2\right)
\arctan\left({1\over s}\right)\right]
\;.
\eqa
Thus $\alpha_2=0$ for $\mu_I^c={1\over2}m_{\pi}$ and we find $\alpha_4>0$
for all values of $m_{\sigma}$, specifically $\alpha_4=0.0514$
for $m_{\sigma}=600$ MeV and $m_{\pi}=140$ MeV.
We therefore conclude that the transition from the vacuum to the 
pion-condensed phase is second order and the onset is for
$\mu_I^c={1\over2}m_{\pi}$.
This results holds for all $\mu<m_q-{1\over2}m_{\pi}$ when the finite-density
part of the potential becomes nonzero, as seen in Fig. \ref{pionphase}.
\end{widetext}


\bibliography{refs}{}

\begin{thebibliography}{99}


\bibitem{raja}
K. Rajagopal and F. Wilczek, 
At the frontier of particle physics, Vol. 3
(World Scientific, Singapore, p 2061) (2001).
\bibitem{alford}
M. G. Alford, A. Schmitt, K. Rajagopal, and T. Sch\"afer,
Rev. Mod. Phys. {\bf 80}, 1455 (2008).

\bibitem{fukurev}
K. Fukushima and  T. Hatsuda,
Rept. Prog. Phys. {\bf 74}, 014001 (2011).

\bibitem{ferrer1}
E. J. Ferrer, V. de la Incera, and C. Manuel
Phys. Rev. Lett. {\bf 95}, 152002 (2005);
Nucl. Phys. B {\bf 747}, 88 (2006).

\bibitem{harmen}
K. Fukushima  and H. J. Warringa, Phys.Rev. Lett. {\bf 100}, 032007 (2008).
\bibitem{nor1}
J. L. Noronha  and I. A. Shovkovy,
Phys. Rev. D {\bf 76}, 105030  (2007);
Phys. Rev. D {\bf 86}, 049901  (2012).

\bibitem{rob1}
T. Kojo, Y. Hidaka, L. McLerran, and R. D. Pisarski, 
Nucl. Phys. A {\bf 843}, 37 (2010); ibid {\bf 875}, 94, (2012).

\bibitem{rob2}
T. Kojo, R. D. Pisarski, and A. M. Tsvelik, Phys. Rev. D {\bf 82}, 074015 
(2010).




\bibitem{casal} 	
R. Anglani, R. Casalbuoni, M. Ciminale, N. Ippolito, R. Gatto, 
M. Mannarelli, and M. Ruggieri, Rev. Mod. Phys. {\bf 86}, 509  (2014).


\bibitem{buballarev}
M. Buballa and S. Carignano, Prog. Part. Nucl. Phys. {\bf 81}, 39 (2015).



\bibitem{kogut1}
J. B. Kogut and D. K. Sinclair,
Phys. Rev. D {\bf 66}, 014508 (2002).
\bibitem{kogut2}
J. B. Kogut and D. K. Sinclair,
Phys. Rev D {\bf 66} 034505  (2002).


\bibitem{gergy1} 	
B. B. Brandt and G. Endrodi,
PoS LATTICE 2016, 039 (2016).

\bibitem{gergy2} 	
B. B. Brandt, G. Endrodi, and S. Schmalzbauer,
e-Print: arXiv:1709.10487 [hep-lat] 

\bibitem{gergy3} 	
B. B. Brandt, G. Endrodi, and S. Schmalzbauer,
e-Print: arXiv:1712.08190.

\bibitem{son}
D. T. Son and M. A. Stephanov,
Phys. Rev. Lett. {\bf 86}, 592 (2001).

\bibitem{kim}
K. Splittorff, D. T. Son, M. A. Stephanov,
Phys. Rev. D {\bf 64}, 016003 (2001).

\bibitem{loewe}
M. Loewe and C. Villavicencio,
Phys. Rev. D {\bf 67}, 074034 (2003).

\bibitem{fragaiso}
E. S. Fraga,  L. F. Palhares and C. Villavicencio,
Phys. Rev. D {\bf 79},  014021 (2009).


\bibitem{carigchpt}
S. Carignano, L. Lepori, A. Mammarella, M. Mannarelli and G. Pagliaroli,
Eur. Phys. J. A {\bf 53}, 35 (2017).

\bibitem{restoublan}
D. Toublan and J. B. Kogut,
Phys. Lett. B {\bf 605}, 129 (2005).

\bibitem{random}
B. Klein, D. Toublan and J. J. M. Verbaarschot,
Phys. Rev. D {\bf 68}, 014009 (2003).



\bibitem{2fbuballa}
M. Frank, M. Buballa, and M. Oertel,
Phys. Lett. B {\bf 562}, 221 (2003).

\bibitem{toublannjl}
D. Toublan, and J. B. Kogut, Phys. Lett. B {\bf 564}, 212 (2003).

\bibitem{bar2f}
A. Barducci, R. Casalbuoni, G. Pettini, and L. Ravagli
Phys. Rev. D {\bf 69}, 096004 (2004).

\bibitem{he2f}
L. He, and P.-F. Zhuang,
Phys. Lett. B {\bf 615}, 93 (2005).

\bibitem{heman2}
L. He, M. Jin and P.-F. Zhuang,
Phys. Rev. D {\bf 71},  116001, (2005).

\bibitem{heman} 	
L. He, M. Jin, and P.-F. Zhuang, 
Phys. Rev. D {\bf 74}, 036005 (2006).

\bibitem{ebert1}	
D. Ebert and K. G. Klimenko,
J. Phys. G {\bf 32},  599 (2006).

\bibitem{ebert2}	
D. Ebert and K. G. Klimenko,
Eur. Phys. J. C {\bf 46}, 771 (2006).

\bibitem{lars} 	
J. O. Andersen and L. Kyllingstad,
J. Phys. G {\bf 37}, 015003 (2009).

\bibitem{2fabuki}
H. Abuki, R. Anglani, R. Gatto, M. Pellicoro, and M. Ruggieri,
Phys. Rev. D {\bf 79}, 034032 (2009).

\bibitem{heman3}
C.-F. Mu, L. He, and Y. Liu, Phys. Rev. D {\bf 82}, 056006 (2010).

\bibitem{he3f}
T. Xia, L. He and P. Zhuang,
Phys. Rev. D {\bf 88}, 056013 (2013).






\bibitem{lorenz}
K. Kamikado, N. Strodthoff, L. von Smekal, and J. Wambach
Phys. Lett. B {\bf 718}, 1044 (2013).

\bibitem{ueda}
H. Ueda, T. Z. Nakano, A. Ohnishi, M. Ruggieri, and K. Sumiyoshi,
Phys. Rev. D {\bf 88}, 074006 (2013).


\bibitem{qmstiele}
R. Stiele, E. S. Fraga and J. Schaffner-Bielich,
Phys. Lett. B {\bf 729}, 72 (2014).
 	
\bibitem{graf}
T. Graf, J. Schaffner-Bielich, and E. S. Fraga,
Phys.Rev. D {\bf 93},085030 (2016).

\bibitem{htlpt}
J. O. Andersen, N. Haque, M. G. Mustafa, and M. Strickland,
Phys. Rev. D {\bf 93}, 054045 (2016).





\bibitem{nakano}
E. Nakano and T. Tatsumi, 
Phys. Rev. D {\bf 71}, 114006 (2005).

\bibitem{nick1}
D. Nickel,
Phys. Rev. {\bf D} 80, 074025 (2009);
D. Nickel,
Phys. Rev. Lett. {\bf 103}, 072301 (2009).


\bibitem{balli}
S. Carignano, D. Nickel, and M. Buballa,
Phys. Rev. D {\bf 82}, 054009 (2010).


\bibitem{bubsc}
S. Carignano, M. Buballa, and B.-J. Schaefer,
Phys. Rev.  D {\bf 90}, 014033 (2014).









\bibitem{1p1klim}
N. V. Gubina, K. G. Klimenko, S. G. Kurbanov, and V. Ch. Zhukovsky,
Phys. Rev. D {\bf 86},  085011 (2012).

\bibitem{jens1+1}
P. Adhikari and J. O. Andersen,
Phys. Rev. D {\bf 95}, 054020 (2017).

\bibitem{khunja}
T. G. Khunjua, K. G. Klimenko, R. N. Zhokhov, V. C. Zhukovsky,
Phys. Rev. D {\bf 95}, 105010  (2017).

\bibitem{dautry} F. Dautry and E. M. Nyman, Nucl. Phys. A {\bf 319}, 323 (1979).


\bibitem{cohen}
T. D. Cohen, Phys. Rev. Lett. {\bf 91}, 222001 (2003).


\bibitem{crew}
P. Adhikari, J. O. Andersen, and P. Kneschke, 
Phys. Rev.  D {\bf 96}, 016013 (2017).


\bibitem{sir1}
A. Sirlin, Phys. Rev. D {\bf 22}, 971 (1980).
\bibitem{sir2}
A. Sirlin, Phys. Rev. D {\bf 29}, 89 (1984).
\bibitem{hollik}
M. Bohm, H. Spiesberger, and W. Hollik, Fortsch. Phys. {\bf 34}, 687 (1986).











\end{thebibliography}
\bibliographystyle{apsrmp4-1}
\end{document}